\documentclass[nojss]{jss}\usepackage[]{graphicx}\usepackage[]{color}
\makeatletter
\def\maxwidth{ %
  \ifdim\Gin@nat@width>\linewidth
    \linewidth
  \else
    \Gin@nat@width
  \fi
}
\makeatother

\definecolor{fgcolor}{rgb}{0.345, 0.345, 0.345}
\newcommand{\hlnum}[1]{\textcolor[rgb]{0.686,0.059,0.569}{#1}}%
\newcommand{\hlstr}[1]{\textcolor[rgb]{0.192,0.494,0.8}{#1}}%
\newcommand{\hlopt}[1]{\textcolor[rgb]{0,0,0}{#1}}%
\newcommand{\hlstd}[1]{\textcolor[rgb]{0.345,0.345,0.345}{#1}}%
\newcommand{\hlkwb}[1]{\textcolor[rgb]{0.69,0.353,0.396}{#1}}%
\newcommand{\hlkwc}[1]{\textcolor[rgb]{0.333,0.667,0.333}{#1}}%
\newcommand{\hlkwd}[1]{\textcolor[rgb]{0.737,0.353,0.396}{\textbf{#1}}}%

\usepackage{framed}
\makeatletter
\newenvironment{kframe}{%
 \def\at@end@of@kframe{}%
 \ifinner\ifhmode%
  \def\at@end@of@kframe{\end{minipage}}%
  \begin{minipage}{\columnwidth}%
 \fi\fi%
 \def\FrameCommand##1{\hskip\@totalleftmargin \hskip-\fboxsep
 \colorbox{shadecolor}{##1}\hskip-\fboxsep
     \hskip-\linewidth \hskip-\@totalleftmargin \hskip\columnwidth}%
 \MakeFramed {\advance\hsize-\width
   \@totalleftmargin\z@ \linewidth\hsize
   \@setminipage}}%
 {\par\unskip\endMakeFramed%
 \at@end@of@kframe}
\makeatother

\definecolor{shadecolor}{rgb}{.97, .97, .97}
\definecolor{messagecolor}{rgb}{0, 0, 0}
\definecolor{warningcolor}{rgb}{1, 0, 1}
\definecolor{errorcolor}{rgb}{1, 0, 0}
\newenvironment{knitrout}{}{} 

\usepackage{alltt}

\usepackage{thumbpdf,lmodern}

\usepackage{framed}

\usepackage{amsmath}
\usepackage{amssymb}
\usepackage{amsfonts}
\usepackage{subfigure}
\usepackage{booktabs}
\usepackage{caption}

\newcommand{\bB}{\textbf{B}}

\newcommand{\bC}{\textbf{C}}

\newcommand{\bI}{\textbf{I}}

\newcommand{\bM}{\textbf{M}}

\newcommand{\bs}{\textbf{s}}

\newcommand{\bV}{\textbf{V}}
\newcommand{\bw}{\textbf{w}}

\newcommand{\bx}{\textbf{x}}
\newcommand{\bX}{\textbf{X}}
\newcommand{\by}{\textbf{y}}
\newcommand{\bY}{\textbf{Y}}

\newcommand{\bR}{\textbf{R}}

\newcommand{\mbs}[1]{\boldsymbol{#1}}

\newcommand{\bSigma}{\mbs{\Sigma}}

\newcommand{\bphi}{\mbs{\phi}}
\newcommand{\bmu}{\mbs{\mu}}
\newcommand{\bbeta}{\mbs{\beta}}

\newcommand{\bepsilon}{{\mbs{\varepsilon}}}

\newcommand{\btheta}{{\mbs{\theta}}}

\newcommand{\ben}{\begin{equation*}}
\newcommand{\een}{\end{equation*}}
\newcommand{\bean}{\begin{eqnarray*}}
\newcommand{\eean}{\end{eqnarray*}}
\newcommand{\bsm}{\begin{smallmatrix}}
\newcommand{\esm}{\end{smallmatrix}}
\newcommand{\bmat}{\begin{matrix}}
\newcommand{\emat}{\end{matrix}}

\newcommand{\bzero}{\textbf{0}}

\newcommand{\given}{\,|\,}
\newcommand{\taus}{\tau^2}
\newcommand{\sigs}{\sigma^2}
\newcommand{\eps}{\epsilon}
\newcommand{\calD}{\mathcal D}
\newcommand{\calR}{\mathcal R}
\newcommand{\tC}{\widetilde \bC}
\newcommand{\tM}{\widetilde \bM}
\newcommand{\tSig}{\widetilde \bSigma}

\newcommand{\bomega}{ \mbox{\boldmath $ \omega $} }
\newcommand{\bin}{\mbox{Binomial}}

\newcommand{\ind}{\overset{ind}{\sim}}
\newcommand{\tR}{\widetilde \bR}

\author{Andrew O. Finley\\Michigan State University
  \And Abhirup Datta\\Johns Hopkins University
   \And Sudipto Banerjee\\University of California,\\Los Angeles}
\Plainauthor{Andrew O. Finley, Abhirup Datta, Sudipto Banerjee}

\title{\pkg{spNNGP} \proglang{R} package for Nearest Neighbor Gaussian Process models}
\Plaintitle{spNNGP R package for Nearest Neighbor Gaussian Process models}
\Shorttitle{\pkg{spNNGP} Nearest Neighbor Gaussian Process models}

\Abstract{
  This paper describes and illustrates functionality of the \pkg{spNNGP} \proglang{R} \citep{R} package. The package provides a suite of spatial regression models for Gaussian and non-Gaussian point-referenced outcomes that are spatially indexed. The package implements several Markov chain Monte Carlo (MCMC) and MCMC-free Nearest Neighbor Gaussian Process (NNGP) models for inference about large spatial data. Non-Gaussian outcomes are modeled using a NNGP P{\'o}lya-Gamma latent variable. \proglang{OpenMP} parallelization options are provided to take advantage of multiprocessor systems. Package features are illustrated using simulated and real data sets. 
}

\Keywords{MCMC, Nearest Neighbor Gaussian Process, kriging, \proglang{R}}
\Plainkeywords{MCMC, Nearest Neighbor Gaussian Process, kriging, R}

\Address{
  Andrew Finley\\
  Department of Forestry\\
  Michigan State University\\
  Natural Resources Building\\
  480 Wilson Road, Room 126\\
  East Lansing, MI 48824-6402\\
  E-mail: \email{finleya@msu.edu}\\
  URL: \url{https://www.finley-lab.com}\\
  \\
  Abhirup Datta\\
  Department of Biostatistics\\
  Johns Hopkins Bloomberg\\
  School of Public Health\\
  615 N Wolfe Street, Room E3640\\
  Baltimore, MD 21205\\
  E-mail: \email{abhidatta@jhu.edu}\\
  URL: \url{http://abhidatta.com}\\
    \\
  Sudipto Banerjee\\
  Fielding School of Public Health\\
  University of California, Los Angeles\\
  650 Charles E. Young Dr. South\\
  Los Angeles, CA 90095-1772\\
  E-mail: \email{sudipto@ucla.edu}\\
  URL: \url{https://ph.ucla.edu/faculty/banerjee}
}
\IfFileExists{upquote.sty}{\usepackage{upquote}}{}
\IfFileExists{upquote.sty}{\usepackage{upquote}}{}
\begin{document}

\section{Introduction} \label{sec:intro}

This paper introduces the \pkg{spNNGP} \proglang{R} \citep{R} package that provides a suite of spatial regression models for Gaussian and non-Gaussian univariate outcomes observed at point-referenced two-dimensional locations. There are, by now, many \proglang{R} packages that provide similar basic functionality. A recent read of the ``Analysis of Spatial Data'' CRAN Task View \citep{CRANSP} yielded $\sim$46 packages listed for geostatistical analysis---and this is not an exhaustive accounting of packages available for such analyses. Our software design focus and unique contribution is to provide Bayesian models and associated diagnostic and prediction functions capable of handling data sets with a large number of locations via the Nearest Neighbor Gaussian Process \citep[NNGP;][]{nngp}. Specifically, functions in \pkg{spNNGP} implement recent methodological and algorithmic developments presented in \cite{finley2019efficient}. 

There have been many recent methodological developments within the large spatial data literature that aim to deliver massively scalable spatial processes. \cite{sunligenton11} and \cite{banerjee2017} provide background and discussion of current work in this area. A recent contribution by \cite{heaton2019case} is particularly useful as it provides an overview of modeling approaches for large spatial data that are under active software development, and a comparison of these approaches based on the analysis of a common dataset in the form of a ``friendly competition.'' In addition to NNGP models, the comparison presented by \cite{heaton2019case} considered covariance tapering via the \pkg{spam} package \citep{Furr:Sain:10, spam}, gap-filling via \pkg{gapfill} \citep{gapfill}, metakriging \citep{GuhBan16}, spatial partitioning \citep{sang2011covariance, barbian2017spatial}, fixed rank kriging via \pkg{FRK} \citep{Cressie_2008,zammit2017frk}, multiresolution approximation \citep{katzfussmultires}, stochastic partial differential equations via \pkg{INLA} \citep{rinla}, lattice kriging via \pkg{LatticeKrig} \citep{nychka2015multiresolution}, local approximate Gaussian processes via \pkg{laGP} \citep{gram14, gramacy2016laGP}, and reduced rank predictive processes \citep{BGFS2008, finley2009} via \pkg{spBayes} \citep{FBG15}. The comparison was based on out-of-sampled predictive performance and, to a lesser extent, computing time for a moderately sized simulated and real dataset comprising 105,569 observations. Comparisons showed NNGP models yielded highly competitive predictive performance and computation time. More recently, \cite{Risser2019} developed the \pkg{BayesNSGP} package for nonstationary Gaussian process modeling with options to use NNGPs for large data settings. In a frequentist setup, fast maximum likelihood-based parameter estimation and predictions using nearest neighbor approximations to the Gaussian Process likelihood are available in the \pkg{GpGp} \citep{guinness2018permutation} and \pkg{BRISC} \citep{briscpkg} packages on CRAN. The latter also offers inference on the spatial covariance parameters using a fast spatial bootstrap \citep{brisc}. While most of the software noted above exploit sparsity in the spatial covariance or precision matrix, or pursue low-rank approximations, the \pkg{ExaGeoStat} package \citep{Abdulah18} tackles decomposition of the full dense spatial covariance matrix head-on using high performance linear algebra libraries associated with various leading edge parallel architectures.

Our contribution here is many fold. We propose a novel extension for analyzing spatially correlated non-Gaussian (binary) responses using a NNGP spatial generalized linear mixed model (GLMM) by using a P{\'o}lya-Gamma prior \citep{polson2013bayesian} for the regression coefficients, which leads to an efficient data-augmented Gibbs sampler. This is, to our knowledge, the first sampler for Binomial spatial-GLMM that ensures closed form Gibbs updates for most parameters using linear time and storage. We discuss model comparison and model adequacy and show how different variants of the NNGP models entail fundamentally different implementation and interpretation of model comparison metrics. This is an important pragmatic consideration for practitioners choosing among different NNGP models. Finally, we provide detailed documentation and exposition of the user-friendly \pkg{spNNGP} \proglang{R} package that: 1) implements NNGP model fitting algorithms for Gaussian response spatial data presented in \cite{nngp} and \cite{finley2019efficient}; 2) provides NNGP models for the non-Gaussian spatial response via the P{\'o}lya-Gamma data-augmented sampler; 3) offers support functions for NNGP model fit diagnostics, summary, and prediction. We discuss code-optimization aspects that allows \pkg{spNNGP} to handle very large data sets by being judicious with memory use, taking advantage of properties of the NNGP dependence scheme to efficiently store and retrieve neighbor information, and applying parallel processing where advantageous. Core model fitting and prediction functions in \pkg{spNNGP} have achieved unprecedented scalability, delivering inference for data sets in the hundreds of millions of spatial locations. 

The remainder of this article proceeds as follows. Section~\ref{sec:gp} provides a brief overview of Gaussian process models followed by specifics about the NNGP models in Section~\ref{sec:nngp}. In Section \ref{sec:pg} we propose extensions for Binomial spatial data and outline the P{\'o}lya-Gamma data-augmented Gibbs sampler. Section \ref{sec:rep} discusses appropriate model comparison and adequacy measures for the different NNGP models. Section~\ref{sec:features} gives a brief description of the code underlying \pkg{spNNGP} and some software development specifics for large data sets. This is followed by analysis of simulated and real data sets in Section~\ref{sec:illustrations} meant to provide a practical tour of some of the package's features. Finally, Section~\ref{sec:summary} provides a brief summary with an eye toward future development.   

\section{Models and software} \label{sec:models}

\subsection{Review of Gaussian Process models for spatial data}\label{sec:gp}
The standard geostatistical paradigm envisions each data unit as a  triplet $(\bs_i,\bx(\bs_i),y(\bs_i))$, where $\bs_i$ denotes the geographical location of the measurements, $\bx(\bs_i)$ is the $p \times 1$ vector of covariates, and $y(\bs_i)$ is the response of interest, for $i=1,2,\ldots,n$. If the covariates fail to account for all of the structured variation observed on the response, a spatial linear mixed effects model for analyzing the data is specified as
\begin{equation}\label{eq:lmm}
y(\bs_i) = \bx(\bs_i)^\top\bbeta + w(\bs_i) + \eps(\bs_i),
\end{equation}
where $\bbeta$ is the $p \times 1$ vector of regression coefficients, $\epsilon(\bs_i)$ is the random noise, customarily modeled as independent and identically distributed (iid) observations from $N(0,\taus)$, and $w(\bs_i)$ is the location-specific spatial random effect. Typically, the spatial random effects are assumed to be smooth across space, i.e., $w(\bs_i)$ can be conceived of as realizations of a smooth latent surface $\{w(\bs) \given s \in \calD\}$, where $\calD$ denotes the geographical domain of interest. Gaussian Processes (GP) are widely used in machine learning for modeling smooth functions on the many-dimensional covariate domain \citep{rasmussen2003gaussian}. For spatial regression, the mean function of the covariates is often specified parsimoniously via a linear model as in (\ref{eq:lmm}) and GPs are typically used to model the latent surface $w(\bs)$ on the two- or three-dimensional physical domain. A GP model for the spatial surface $\{w(\bs)\}$, denoted by $w(\bs) \sim GP(0,C(\cdot,\cdot \given \btheta))$ where $C(\cdot,\cdot \given \btheta)$ is a covariance function, implies that for any finite set of locations $\bs_1, \ldots, \bs_n$, the vector of random effects $\bw=(w(\bs_1), \ldots, w(\bs_n))^\top $ follows a zero-mean multivariate Gaussian distribution with covariance matrix $\bC(\btheta)=\bC=(c_{ij})$ where $c_{ij}=C(\bs_i,\bs_j \given \btheta)$. The parametric covariance function $C(\cdot,\cdot \given \btheta)$ is often selected from the Mat\'ern family of functions \citep{stein2012interpolation}, popular due to its versatility to model surfaces with varying degrees of smoothness. 

The mixed effects model for the response and the GP model for the random effects can be combined into the hierarchical model
\begin{equation}\label{eq:gphier}
\by \sim N(\bX\bbeta + \bw, \taus \bI),\; \bw \sim N(\bzero, \bC(\btheta)), 
\end{equation}
where $\by$ is the vector formed by stacking the $y(\bs_i)$'s, and $\bX$ is the corresponding $n \times p$ matrix of covariates. Equivalently, one can integrate out $\bw$ from (\ref{eq:gphier}) and write 
\begin{equation}\label{eq:gpmar}
\by \sim N(\bX\bbeta , \bC(\btheta) + \taus \bI).
\end{equation}
Parameter estimation in a frequentist paradigm maximizes the likelihood from (\ref{eq:gpmar}) with respect to $\bbeta$, $\taus$ and $\btheta$, while in a Bayesian framework, priors are assigned to these parameters, and one can use either the hierarchical model (\ref{eq:gphier}) or the marginal model (\ref{eq:gpmar}) to obtain posterior inference via Markov chain Monte Carlo (MCMC). 

\subsection{Nearest Neighbor Gaussian Processes for large spatial data}\label{sec:nngp}
Gaussian Processes encounter computational roadblocks when data is observed at a large number of locations. Both the joint likelihood from the hierarchical model (\ref{eq:gphier}) or the data likelihood from the marginalized model (\ref{eq:gpmar}) involves a multivariate Gaussian distribution with a dense $n \times n$ covariance matrix ($\bC(\btheta)$ and $\bC(\theta) + \taus \bI$, respectively). This task involves $O(n^2)$ storage, and $O(n^3)$ computations (floating point operations or FLOPs), which is infeasible when $n$ is large. 

One of the scalable solutions is replacing the GP prior for the spatial random effects with a {\em Nearest Neighbor Gaussian Process} prior \citep{nngp}. Let  $\calR=\{\bs_1,\ldots,\bs_n\}$ be any fixed finite set of locations in the spatial domain $\calD$, and $\bw_\calR=(w(\bs_1),\ldots,w(\bs_n))^\top$. For any location in $\calD$, define neighbor sets as follows
\begin{equation}\label{eq:nei}
\begin{aligned}
    N(\bs_1) &= \{\} \mbox{ (empty set)},\\
    N(\bs_i) &= \min(m,i-1) \mbox{ nearest neighbors of } \bs_i \mbox{ in } \bs_1,\ldots,\bs_{i-1}, \mbox{ for } i=2,\ldots,n,\\
    N(\bs) &= m \mbox{ nearest neighbors of } \bs \mbox{ in } \calR.
\end{aligned}
\end{equation}
Then NNGP is  specified as 
\begin{equation}\label{eq:nngp}
\bw_\calR \sim \prod_{i=1}^n p(w_i \given \bw(N(\bs_i));\;\; w(\bs) \given \bw_\calR \ind  p(w(\bs) \given \bw(N(\bs)) \mbox{ for all } \bs \in \calD \setminus \calR.
\end{equation}
In practice, $\calR$ is chosen to be the data-locations and (\ref{eq:nngp}) 
yields the NNGP approximation to the distribution of the spatial effects. If (\ref{eq:nngp}) is applied to the response vector, then we obtain the likelihood approximation. This is motivated from the likelihood approximation ideas in \cite{ve88} and \cite{stein2004}. 
 \cite{nngp} showed that the above construction endows a multivariate Gaussian distribution on $\bw=\bw_\calR$
\begin{equation}\label{eq:nngpw}
\bw \sim N(\bzero, \tC(\btheta)) \;,
\end{equation}
where $\tC(\btheta)^{-1}$, i.e., the inverse of the NNGP covariance matrix, is sparse. The extension to arbitrary locations in the bottom-row of (\ref{eq:nngp}) is based on kriging (conditional) distributions using only $m$-nearest neighbors from the data-locations. Nearest-neighbor kriging  have been explored in \cite{emery2009kriging} in the context of point-predictions as opposed to the full predictive (conditional) distributions specified for NNGP. The NNGP is itself a valid stochastic (Gaussian) process on the entire domain $\calD$---a chief distinction from frequentist likelihood approximations or prediction equations.   
If probabilistic modeling of a spatial surface (observed or latent) is of primary concern, then the NNGP is a legitimate candidate and need not be viewed as an approximation of the parent GP from which it is derived. Only when inference on the spatial covariance parameters are of interest, NNGP needs to be perceived as an approximation, as these parameters describe attributes of the parent GP. The NNGP covariance function $\widetilde C$ is constructed from the original covariance function $C$ and ensures that the matrix $\tC(\btheta)^{-1}$ is sparse and, consequently, the likelihood (\ref{eq:nngpw}) can be evaluated using only $O(n)$ storage. This makes NNGP a scalable replacement for the full GP, while delivering inference almost indistinguishable from the full GP.  NNGP can, in fact, also be looked upon as a special case of a Gaussian Markov Random Field \citep[GMRF;][]{rue2005gaussian} with a specific neighborhood structure defined through the neighbor sets of (\ref{eq:nei}). In practice, however, NNGP delivers inference sufficiently close to traditional Gaussian random fields without requiring mesh-based finite element approximations of SPDE representations of Gaussian random fields (as done in \cite{lindgren11}). 

\subsubsection{Latent NNGP}\label{sec:seq}
The original implementation of the NNGP model proposed in \cite{nngp} used a fully Bayesian hierarchical specification 
\begin{equation}\label{eq:nngpseq}
N(\by \given \bX\beta + \bw, \taus \bI) \times N(\bw \given \bzero, \tC(\btheta)) \times p(\bbeta,\btheta,\taus) 
\end{equation}
for running an MCMC algorithm, where all the parameters $\{\bw,\bbeta,\btheta,\taus\}$ are updated in a Gibbs sampler. Normal priors for $\bbeta$ and inverse-Gamma priors for the variance components ensure that they produce conjugate full conditionals in the Gibbs sampler. The remaining covariance parameters are updated using random-walk Metropolis steps for their respective full conditional distributions. The full conditional distribution for $\bw$ from (\ref{eq:nngpseq}) is 
\begin{equation*}
\bw \given \cdot \sim N(\bB (\by - \bX\bbeta)/\taus), \bB) \mbox{  where } \bB=\tC(\btheta)^{-1} + \bI/\taus\;.
\end{equation*}
Unfortunately, this block update of $\bw$ is not practical. While the full conditional precision matrix $\tC(\btheta)^{-1} + \bI/\taus$ has the same sparsity as $\tC(\btheta)^{-1}$, unlike $\tC(\btheta)^{-1}$ its determinant cannot be calculated in $O(n)$ FLOPs. Instead, \cite{nngp} recommended sequentially updating the full conditionals $w_i \given \cdot$ for $i=1,\ldots,n$ and outlined an algorithm that accomplishes this entire sequence of updates in $O(n)$ FLOPs. We refer to this NNGP model as the {\em latent NNGP}. 

\subsubsection{Response NNGP}\label{sec:res} 
The latent NNGP algorithm involves running an MCMC of dimension $O(n)$ where each of the $n$ parameters $w_i$ are sequentially updated. While this ensures linear scalability of the NNGP model with sample size, it can also imply very slow convergence of the high-dimensional MCMC. Instead of using NNGP for the latent Gaussian process $w(\bs)$, \cite{finley2019efficient} directly considered the marginal Gaussian process for the response, i.e., $\{y(\bs)\} \sim GP(\bx(\bs)^\top\beta,\Sigma(\cdot,\cdot))$ where the marginalized covariance function $\Sigma$ is specified as $\Sigma(\bs_i,\bs_j) = C(\bs_i,\bs_j \given \btheta) + \taus \delta(\bs_i,\bs_j)$, $\delta$ denoting the Kronecker delta. Since the covariance function of an NNGP can be derived from any parent GP, \cite{finley2019efficient} replaced the covariance function $\Sigma$ with its NNGP analogue $\widetilde \Sigma$ yielding the response model 
\begin{equation}\label{eq:nngpres}
\bY \sim N(\bX\bbeta, \tSig), 
\end{equation}
where $\tSig$ is the NNGP covariance matrix derived from $\bSigma = \bC(\btheta) + \taus \bI$. The advantage of this {\em response NNGP} model over the previous latent model is that the dimension of the parameter space is drastically reduced from $O(n)$ to $O(1)$. The lower dimensional NNGP tends to have improved MCMC convergence as shown in \cite{finley2019efficient}. The computational scalability per MCMC iteration of the response model remains the same as in the latent model algorithm, as $\tSig^{-1}$ is sparse requiring $O(n)$ storage and its quadratic form and determinant can be calculated using $O(n)$ FLOPs. 

\subsubsection{MCMC-free inference: Conjugate NNGP model}\label{sec:conj} 
The conjugate NNGP algorithm is an implementation of the response NNGP model which fixes certain spatial covariance parameters leading to exact (MCMC-free) posterior Bayesian inference. Recall that under the full-GP specification, the covariance function for $y(\bs)$ is specified as $\Sigma(\bs_i,\bs_j) = C(\bs_i,\bs_j \given \btheta) + \taus \delta(\bs_i,\bs_j)$. Usually, the covariance functions $C(\cdot, \cdot \given \btheta)$ can be expressed as $\sigs R(\cdot, \cdot \given \bphi)$ where $\sigs$ is the marginal variance, and $R$ is the correlation function parameterized by $\bphi$, i.e., $\btheta=(\sigs,\bphi)$. Rewriting $\taus = \alpha\sigs$, we have $\Sigma(\bs_i,\bs_j) = \sigs (R(\bs_i,\bs_j \given \bphi) + \alpha\; \delta(\bs_i,\bs_j))$. The conjugate NNGP model fixes $\bphi$ and $\alpha$, and generates the NNGP covariance function approximation $\widetilde M(\cdot,\cdot \given \alpha, \bphi)$ of $R(\cdot,\cdot \given \bphi) + \alpha\; \delta(\cdot,\cdot)$. This implies we have the following marginal model
\begin{equation}\label{eq:conj}
\bY \sim N(\bX\bbeta, \sigs \tM)
\end{equation}
where $\tM = \tM(\bphi,\alpha)$ is a known covariance matrix once $\bphi$ and $\alpha$ are fixed. The model in (\ref{eq:conj}) is the standard Bayesian linear model with only unknowns $\bbeta$ and $\sigs$. Using a Normal-Inverse-Gamma prior for ($\bbeta,\sigs$) leads to conjugate Normal-Inverse-Gamma posterior distributions  and hence posterior moments and other summary quantities of $\bbeta$ and $\sigs$ are easily and exactly obtained. Exact posterior predictive distributions for $y(\bs_0)$ at a new location $\bs_0$ are also available. The matrix $\tM^{-1}$ (like $\tSig^{-1}$ for the response model) is sparse ($O(n)$) ensuring all the posterior distributions and moments can be evaluated using $O(n)$ memory and FLOPs. The fixed values of $\bphi$ and $\alpha$ are either chosen based on a variogram or can be selected in a more formal fashion using cross-validation using predictions from the model on hold-out data. We describe an implementation of the cross-validation in more detail in the \emph{MCMC-free inference} subsection of Section \ref{sec:illustrations}. While the cross-validation enforces multiple runs of the conjugate model for a grid of values of $\bphi$ and $\alpha$, these runs can proceed in parallel. Also, $\bphi$ is usually one or two dimensional, hence, there are only 2 or 3 tuning parameters for the model and empirical results in \cite{finley2019efficient} suggest that a crude-resolution grid  often suffices to yield accurate predictive inference. Empirical comparisons in \cite{finley2019efficient} and also in \cite{heaton2019case} suggest that the conjugate NNGP is orders of magnitude faster than the MCMC-based NNGP algorithms and also other competing big-spatial data methods while delivering highly competitive prediction performance.

\subsection{Binomial response}\label{sec:pg}
All current implementations of the NNGP model assume that the response is Gaussian. Here we implement a Bayesian NNGP model for Binomial response. Note that the response and the conjugate NNGP model explicitly rely on Gaussian distributions to derive the marginal distribution for the response $\by$. Such closed form marginal distributions are not available for non-Gaussian responses. However, conceptually we can extend the latent NNGP model of (\ref{eq:nngpseq}) to non-Gaussian settings. Assuming $y(\bs_i) \ind \bin(n_i, \frac 1{1 + \exp(- \psi(\bs_i))})$, where $\psi(\bs_i) = \bx(\bs_i)^\top\bbeta + w(\bs_i)$ with $w(\bs_i)$ modeled as a NNGP, 
the joint likelihood for this model is given by:
\begin{equation}\label{eq:ngnngp}
\prod_{i=1}^n \bin(y(\bs_i) \given n(\bs_i), \frac 1{1 + \exp(- \psi(\bs_i))}) \times N(\bw \given \bzero, \tC(\btheta)) \times p(\bbeta,\btheta) \;.
\end{equation}
Since $\bw$ is given an NNGP prior, this likelihood can still be evaluated using $O(n)$ memory and FLOPs. However, unlike the latent NNGP model for the Gaussian case, where the full conditional distributions $w(\bs_i) \given \cdot$ used in the Gibbs updates were Gaussian distributions, for the non-Gaussian case these full conditionals $w(\bs_i) \given \cdot$ do not belong to any standard family. Alternatively, one can resort to Metropolis-Hastings (MH) random walk updates for each $w(\bs_i)$. Introducing $n$ MH random walk updates in every iteration would exacerbate the slow convergence issues already plaguing the latent NNGP model. 

To eschew the numerous Metropolis random-walk updates, we devise a Gibbs sampler for the latent NNGP model for Binomial responses exploiting the P{\'o}lya-Gamma data-augmented sampler of \cite{polson2013bayesian}. Again, we write $\tC(\btheta) = \sigs \tR(\bphi)$ where $\tR$ is the NNGP approximation of the GP correlation matrix. We then assume a $N(\bmu, \bV)$ prior for $\bbeta$, an $IG(a,b)$ prior for $\sigs$, and $p(\bphi)$ prior for the other covariance parameters $\bphi$. 
Letting $\kappa(\bs_i) = y(\bs_i) - n(\bs_i) / 2$ we introduce the augmented data $\bomega = (\omega(\bs)_1), \ldots, \omega(\bs_n))^\top$. 
We can then write the conditional likelihood
\begin{equation}\label{eq:likpol}
\begin{array}{cc}
p(\bbeta, \btheta, \bw \given \by, \bomega) \propto & \prod_{i = 1}^n \exp(-\frac {\omega(\bs_i)} 2 (\bx(\bs_i)^\top\bbeta + w(\bs_i)  - \kappa(\bs_i)/\omega(\bs_i))^2) \times \\
& \left( \frac 1 {\sigs} \right) ^\frac n2 \exp(- \frac 1{2\sigs} \bw^\top\tR(\bphi)^{-1}\bw) \times \\
& \exp(-\frac 12 (\bbeta - \bmu)^\top\bV^{-1}(\bbeta-\bmu)) \times  \left( \frac 1 {\sigs} \right) ^{a+1} \exp(\frac b \sigs) \times p(\bphi). 
\end{array}
\end{equation}
Using the augmented data $\bomega$, this likelihood for the Binomial  mixed linear model is similar to the usual likelihood from a spatial mixed linear model with NNGP using responses $y(\bs_i)^* = \kappa(\bs_i)/\omega(\bs_i)$ and heteroskedastic variances $\taus(\bs_i)=1/\omega(\bs_i)$. 

An immediate consequence is that the updates for $\bbeta$ and $\bw$ are analogous to the updates in the NNGP with $y(\bs_i)$ replaced with $y(\bs_i)^*$ and the constant nugget $\taus$ replaced by $\taus(\bs_i)$. Note that for $y(\bs_i)^*$ and $\taus(\bs_i)$ are functions of the augmented data $\bomega$ and will change in every iteration. 
Other than that, the updates remain exactly the same as in the latent model. 

The only additional update we need to do for non-Gaussian responses is that of the $\omega(\bs_i)$'s. We update these using the full conditionals
\[\omega(\bs_i) \sim PG(n(\bs_i), \bx(\bs_i)^\top\bbeta + w(\bs_i)) \], where 
$PG(b,z)$ denotes the P{\'o}lya-Gamma random variable with shape parameter $b$ and tilting parameter $z$ \citep[see][for more details about this distribution]{polson2013bayesian}. 

This completes a Gibbs sampler for spatial GLMM with Binomial responses, where, like the Gaussian case, every parameter except the covariance parameters ($\bphi$) have closed form full conditionals. The memory requirements also remain $O(n)$ thereby ensuring that the Binomial NNGP model is a viable candidate for analysis of massive non-Gaussian spatial data. 

\subsection{Fitted values, replicates, and predictions}\label{sec:rep}
Subsequent to convergence of the MCMC for the latent or response models, we can use the posterior samples of the parameters to generate various quantities of interest like fitted values and replicates at each data location, and predictions at new locations. For the response model, samples for the fitted value of the mean at the $i^{th}$ location is simply given by $\{\bx(\bs_i)^\top\bbeta^{(l)}\}$ where $\bbeta^{(l)}$ denotes the $l^{th}$ post burn-in sample. Similarly, for the latent NNGP, the posterior distribution of the fitted values is specified by the samples $\{\bx(\bs_i)^\top\bbeta^{(l)}+w(\bs_i)^{(l)}\}$. For the MCMC-free conjugate NNGP, let $\hat\phi$ and $\hat\alpha$, respectively, denote the chosen values of $\phi$ and $\alpha$ based on cross-validation using RMSE of predictions on hold-out (test) data. Then we can express the posterior predictive distribution as a t-distribution with mean $\bx(\bs_i)^\top E_{\hat\phi,\hat\alpha}(\bbeta \given \by)$ where $E_{\hat\phi,\hat\alpha}(\bbeta \given \by)$ denotes the posterior mean of $\bbeta$ at $\phi=\hat\phi$ and $\alpha=\hat\alpha$. The closed-form expressions for $E_{\hat\phi,\hat\alpha}(\bbeta \given \by)$ and the scale parameter for the t-distribution for the fitted value are available in Algorithm 5 of \cite{finley2019efficient}. 

Following \cite{gelman2013} we consider generating replicate data at the observed data locations. This is often used for model checking and model adequacy evaluations and is based upon simulating each data point from its posterior predictive distribution. We note that while both the latent model (\ref{eq:gphier}) and the response model (\ref{eq:gpmar}) lead to the same marginal distribution for $\by$, the two models differ fundamentally in their definition for replicate data. For the hierarchical model, conditional on $\bbeta$, $\btheta$ and $\bw$'s, the $y(\bs_i)$'s are independent with iid  $N(0,\taus)$ distributed error terms. 
Hence, for each post-burn-in sample of the parameters, replicates at each data location can be generated as $\{y(\bs_i)^{(l)} \sim N(\bx(\bs_i)^\top\bbeta^{(l)} + w(\bs_i)^{(l)}, {\taus}^{(l)})\}$. For the response model, however, we can think of $\by$ as a single multivariate observation from $N(\bX\beta, \bSigma)$. Things are further complicated when switching to the NNGP covariance matrix $\tSig$ as, unlike $\bSigma$, we cannot express $\tSig$ as the sum of a purely spatial covariance matrix and a diagonal matrix of error variances. Hence, replicates for the marginalized model are basically new multivariate draws from $N(\bX\bbeta^{(l)}, \tSig(\btheta^{(l)}))$. Unlike the replicates from the hierarchical model, which are expected to exhibit strong spatial alignment with the observed data owing to the correlation induced through the use of the common $\bw$'s, the replicates for the response model are only correlated with the observed data through $\bbeta$ and $\btheta$ and hence are not expected to have similar spatial contours as the observed data. For the conjugate NNGP, which also relies on the marginalized model, the replicates are draws from $N(\bX\bbeta^{(l)}_{\hat\bphi,\hat\alpha}, {\sigs}^{(l)}_{\hat\bphi,\hat\alpha} \tM(\hat\bphi,\hat\alpha))$ with $\{(\bbeta^{(l)}_{\hat\bphi,\hat\alpha},{\sigs}^{(l)}_{\hat\bphi,\hat\alpha})\}$ denoting the Normal-Inverse-Gamma posterior samples of $\bbeta$ and $\sigs$ at the chosen value of $\hat\phi$ and $\hat\alpha$. 

Finally, we turn our attention to predictions at a new location $\bs$. For the latent model, predictions are also generated hierarchically by first generating samples of $w(\bs)^{(l)} \given \bw(N(\bs))^{(l)}, \btheta^{(l)}$. This conditional distribution is Gaussian with mean and variance being the kriging mean and variance at $\bs$ based on a set of nearest neighbors $N(\bs)$. The exact expressions for these quantities are specified in Algorithm 2 of \cite{finley2019efficient}. Subsequent to generating samples of $w(\bs)$, we generate replicates for the response as $y(\bs)^{(l)} \given w(\bs)^{(l)}, \bbeta^{(l)}, {\taus}^{(l)} \sim N(\bx(\bs)^\top\bbeta^{(l)} + w(\bs)^{(l)}, {\taus}^{(l)})$. For the response model, the prediction algorithm is presented in Algorithm 4 of \cite{finley2019efficient} and involves directly generating samples of $y(\bs)^{(l)} \given y(N(\bs)), \bbeta^{(l)}, \btheta^{(l)}$. This is once again a conditional normal distribution equivalent to kriging at $\bs$ given its neighbors $N(\bs)$ for the response process $y(\cdot)$. For the conjugate model, exact predictive distributions are available as $y(\bs) \given$data following a $t$-distribution with location and scale parameters provided in Algorithm 5 of \cite{finley2019efficient}. Finally, for Binomial responses, predictions follow the same strategy as the latent model for Gaussian responses as both models rely on the hierarchical latent variable formulation. We initially generate samples of $w(\bs)^{(l)} \given \bw(N(\bs))^{(l)}, \btheta^{(l)}$ followed by generating Binomial samples for the response as $y(\bs)^{(l)} \given w(\bs)^{(l)}, \bbeta^{(l)}, {\taus}^{(l)} \sim \bin(n(\bs), (1+exp(-(\bx(\bs)^\top\bbeta^{(l)} + w(\bs)^{(l)})))^{-1})$ where $n(\bs)$ denotes the total count at $\bs$. 

\subsection{Software features}\label{sec:features}

Two model fitting functions \code{spNNGP} and \code{spConjNNGP}, with associated support functions, offer users a set of efficient regression and prediction tools that implement the methods outlined in Sections~\ref{sec:nngp} and \ref{sec:pg}. Following from Section~\ref{sec:nngp}, \code{spNNGP} provides MCMC-based inference for the latent and response spatial regression models. The latent model can be specified for Gaussian and Binomial (via the P{\'o}lya-Gamma distribution Section~\ref{sec:pg}) outcome variables, while the response model only fits Gaussian outcome variables. The highly efficient MCMC-free conjugate NNGP algorithm used for Gaussian outcome variables described in Section~\ref{sec:nngp} is available via the \code{spConjNNGP} function. 

The \code{spNNGP} and \code{spConjNNGP} functions as well as their support functions (\code{fitted} for generating regression fitted and replicated data, \code{spDiag} for computing model diagnostics, and \code{predict} for sampling from posterior predictive distributions) are written in \proglang{C}/\proglang{C++} using \proglang{R}'s foreign language interface and offer parallelization using \proglang{openMP} \citep{openmp98}. 

As noted previously, our aim was to provide software capable of handling data sets with 10s to 100s of millions of locations. This aim was met by minimizing memory requirements, and taking advantage of parallelization where possible. We attempted to minimize memory requirements in several ways. First, we avoid making copies of input data, e.g., the regression design matrix, spatial coordinate matrix, etc. Second, we do not hold any Euclidean distance or subsequent spatial covariance vectors and matrices in memory, but rather compute them when needed. For example, all sampling algorithms require the computation of the spatial covariance vector between each observed location and its set of nearest neighbors, and the among neighbors spatial covariance matrix. For any given observation this storage requirement is not large, i.e., at most a $m$ length vector and upper or lower triangle of an $m\times m$ matrix, where $m$ is the number of neighbors (which is small, e.g., $m=15$). However, when $n$ is large, storing $n$ $m$ length vectors and $m\times m$ matrices is substantial. Therefore, we made the decision to compute these covariance vectors and matrices ``on the fly'' for each observation and for each iteration of the given sampler. To do this efficiently, we find each observation's neighbor set via a fast code book search described in \cite{Ra1993} which is modified to accommodate the ordering. The neighbor search is performed only once at the beginning of the program, and the result is $n$ vectors of at most $m$ integers that record the data row index for each observation's neighbors (i.e., the indexes in the input outcome vector and corresponding design and coordinate matrices). 

The ordering and subsequent selection of $m$-(directed) nearest neighbors can be summarized as a directed acyclic neighbor graph. This neighbor graph can be encoded as a sparse lower-triangular matrix where each row is an observation and the non-zero elements in each row hold the neighbor indexes. Within the software, this matrix is held in Compressed Row Storage (CRS) format to minimize memory use and neighbors' data retrieval time. The CRS format does not store zeros and organizes non-zero matrix elements ordered by row in contiguous memory. This is advantageous in our setting because it makes traversing all observations' neighbor indexes as simple as a single loop over the vector that holds the CRS matrix's non-zero elements. Given the index for a given neighbor, retrieving its spatial coordinates, design matrix, and outcome value is further simplified because all input data are stored in contiguous memory column-major format, which allows for fast retrieval of data using integer indexing. This straightforward looping and use of integer indexing to retrieve neighbor input data facilitates parallelization using an \proglang{openMP} \code{omp for} pragma at several points in the sampling algorithms. 

\begin{figure}[!htp]
  \centering
    \subfigure[]{\includegraphics[trim=0cm 0cm 0cm 0cm,clip,width=7cm]{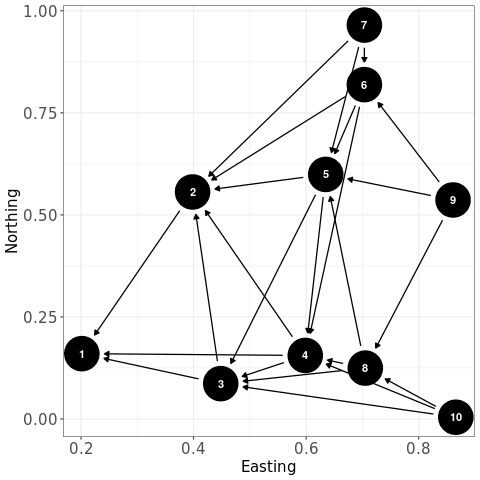}\label{nnGraph}}\hspace{0.75cm}
    \subfigure[]{\includegraphics[trim=0cm 0cm 0cm 0cm,clip,width=7cm]{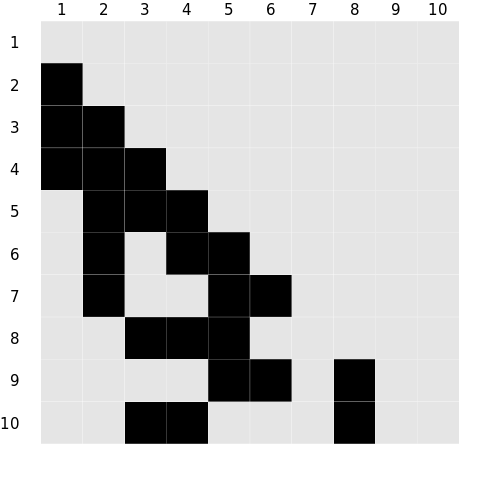}\label{nnMtrix}}
  \caption{\subref{nnGraph} Illustration of a neighbor graph using three neighbors for ten observations with ordering along the easting axis. \subref{nnMtrix} The sparse lower-triangular matrix (black elements are non-zero) corresponding to the graph in \subref{nnGraph} that records the neighbor index (columns) for each observation index (rows).}\label{nnGraphMtrix}
\end{figure}

Figure~\ref{nnGraphMtrix} provides an example of a neighbor graph and the corresponding sparse matrix that we store in CRS format. Following the sampling algorithms in \cite{nngp} and \cite{finley2019efficient}, the response and conjugate models require only each observation's neighbor information; however, the latent model sampling algorithm requires information on which observations have a given observation as a neighbor. Or put another way, we need to know which neighbor sets each observation is a member of. This information is easily accessed by noting the row index of each column's non-zero elements in Figure~\ref{nnMtrix}, e.g., observation 1 is in the neighbor sets for observation 2, 3, and 4. Efficiently accessing and traversing these indexes is done in the software by converting the CRS neighbor index matrix to a Compressed Column Storage (CCS) format, which effectively reorders the needed row indexes in contiguous memory by column.

Storing all vectors and matrices in contiguous memory column-major format also simplifies calls to \proglang{Fortran} Basic Linear Algebra Subprograms (BLAS; \url{www.netlib.org/blas}) and Linear Algebra Package (LAPACK; \url{www.netlib.org/lapack}), which are used for all computationally intensive matrix operations. Hence, additional speed-up can be realized if \proglang{R} calls a threaded implementation of BLAS, e.g., \texttt{openBLAS} \citep{zhang13} or Intel's Math Kernel Library (MKL) \citep{intelMKL}, while working on a multiple processor computer. \cite{finley2019efficient} provide details about where and how each sampling algorithm uses parallelization.

\section{Illustrations}\label{sec:illustrations}

\subsection{Analysis of simulated data}\label{sec:sim-data}

The basic functionality of \code{spNNGP} and \code{spConjNNGP}, along with their support functions, are illustrated using a small $n$=5000 simulated data set with locations distributed at random within a unit square domain. This data set was kept purposely small so the NNGP models could be compared with the full GP model. The Gaussian outcome variable was generated from (\ref{eq:gphier}) with latent $\bw$ centered on zero and covariance matrix elements $c_{ij}=\sigma^2 \exp(-\phi ||\bs_i-\bs_j||)$, where $\phi$=6 and $\sigma^2$=1. The columns in the design matrix comprise an intercept and variable \code{x} which was drawn from a Normal distribution with mean zero and variance 1. The regression coefficients were $\bbeta = (1,-0.1)^\top$ and measurement error $\tau^2$=0.1. An additional $n_0$=2500 observations on a grid were set aside to illustrate prediction.
 
\subsubsection{MCMC-based inference}\label{sec:mcmc_based}

A call to \code{spNNGP} generates samples from parameters' posterior distributions and, optionally, corresponding samples of regression fitted values and replicates via composition sampling as described in Section~\ref{sec:rep}. Similar to \proglang{R}'s \code{lm} function, the desired model is passed to \code{spNNGP} via the \code{formula} argument. The model's outcome vector and design matrix components are then found in a data frame passed to the \code{data} argument or, if \code{data} is not specified, taken from the calling environment. Additionally, the locations corresponding to the observed data are passed as a matrix via the \code{coords} argument. One can also pass \code{coords} a character vector of column names in \code{data}'s data frame.  

Specifying the latent or response model is done via the \code{method} argument. As detailed in \cite{finley2019efficient}, for the latent model (initialized by setting \code{method="latent"}) $\bbeta$, $\bw$, $\sigma^2$, and $\tau^2$ are updated from their respective full conditional distributions via a Gibbs algorithm. The spatial correlation function decay parameter $\phi$ and, if \code{cov.model="matern"} (where \code{cov.model} specifies the desired spatial correlation function), smoothness parameter $\nu$ are updated via a MH algorithm. If \code{method="response"}, only $\bbeta$ has an efficient closed form full conditional Gibbs update and the remaining parameters $\sigma^2$, $\tau^2$, $\phi$, and perhaps $\nu$ are updated via MH. For both models, starting values and prior distributions must be specified for all parameters (with the optional exception of $\bbeta$ and $\bw$) via the \code{starting} and \code{priors} arguments, as illustrated below. Those parameters updated via MH are transformed to have support on the real line so that a Normal proposal distribution can be used. The variance of the parameter specific proposal distribution is controlled via the \code{tuning} argument. Tuning values should be selected to maintain the desired MH acceptance rate (in general we aim for $\sim$25-50 percent, see \cite{Roberts09} and \cite{gelman2013} for guidance). Information about acceptance rate and other model specifics is printed to the console when \code{verbose=TRUE}.

A NNGP model requires two user inputs: ordering of the locations, and number of nearest neighbors $m$ to use. The top-row of Equation (\ref{eq:nngp}) clearly depends on the ordering of the locations in $\calR$. As defined in Equation \ref{eq:nei}, given an ordering, for a data-location corresponding to an observation, the neighbor set (conditioning set) is constructed as the set of  
$m$ nearest neighbors of $\bs$ that conform to the user-specified ordering (i.e., directed nearest neighbors). 
By default, \code{spNNGP} and \code{spConjNNGP} order in increasing value of the first column of the matrix passed to \code{coords}, which, in most settings, produces an excellent approximation to a full GP (see supplemental experiments in \cite{nngp}); however, as demonstrated by \cite{guinness2018permutation} improvements in parameter estimation efficiency can be achieved with different ordering designs. Functions to obtain these alternate orderings are provided in the \pkg{GpGP} \proglang{R} package \citep{guinness2018permutation}.  If the user wishes to try different orderings, they can be passed as an integer index vector via the optional \code{ord} argument. For illustration, we bypass the default ordering in the call to \code{spNNGP} below and order using the sum of locations' $x$ and $y$ coordinate values. We reemphasize that, subsequent to selection of any ordering, NNGP always uses nearest neighbors corresponding to that ordering. Other neighbor selection schemes such as mixture of nearest and farthest neighbors have been considered in related earlier work \citep{stein2004} but is not implemented in spNNGP.  The number of neighbors to consider is controlled by the \code{n.neighbors} argument. As demonstrated in \cite{nngp} 15 neighbors is usually sufficient; however, depending on the data, one can achieve a good approximation to the full GP with as few as five neighbors. If $n$ is large, selecting fewer neighbors can result in substantial decrease in runtime.

Users can choose a slow brute force or fast code book nearest neighbor search algorithm by setting the \code{search.type} argument to \code{"brute"} or \code{"cb"}, respectively. The fast code book search is a modified version of the algorithm detailed in \cite{Ra1993}. If locations do not have identical coordinate values on the axis used for the ordering then \code{"cb"} and \code{"brute"} should produce identical neighbor sets. However, if there are identical coordinate values on the axis used for ordering, then the search algorithms might produce different, but equally valid, neighbor sets. If $n$ is large (e.g., a million or more), constructing the nearest neighbor sets can take a long time even when \code{search.type="cb"}. To save time, the neighbor set can be reused in subsequent model calls if the values passed to \code{coords}, \code{ord}, and \code{n.neighbors} do not change. Setting \code{return.neighbor.info = TRUE} in \code{spNNGP} or \code{spConjNNGP} returns the necessary neighbor information in an object called \code{neighbor.info}. Then passing this object to the optional \code{neighbor.info} argument in subsequent calls to \code{spNNGP} or \code{spConjNNGP} avoids the costly nearest neighbor search. The information in \code{neighbor.info} can also be used if one wishes to plot the neighbor sets (see example code in the \code{spNNGP} manual page).

The call to \code{spNNGP} in the code below generates 2000 MCMC samples using the latent model (with the \code{n.samples} argument specifying the desired number of samples). The \code{n.report} argument defines the sample interval for reporting the MH acceptance rate, which in this case is once every 1000 samples. The \code{fit.rep=TRUE} indicates that we are requesting regression fitted and replicate data be generated via composition sampling (i.e., one-for-one with each MCMC sample). The argument \code{sub.sample} specifies that we only want regression fitted and replicate samples starting at MCMC sample 1000 and for each subsequent sample, i.e., \code{start=1000} (the list passed to \code{sub.sample} can also define \code{end} and \code{thin} for additional control over MCMC chain samples used in the computations).

The computer used to conduct this analysis has multicore processor and \proglang{R} compiled with \proglang{openMP}. Therefore, setting \code{n.omp.threads=4} should decrease runtime, see Section~\ref{sec:runtime} for more details on computing time.

\begin{knitrout}
\definecolor{shadecolor}{rgb}{0.969, 0.969, 0.969}\color{fgcolor}\begin{kframe}
\begin{alltt}
\hlstd{R> }\hlstd{n.samples} \hlkwb{<-} \hlnum{2000}
\hlstd{R> }\hlstd{starting} \hlkwb{<-} \hlkwd{list}\hlstd{(}\hlstr{"phi"}\hlstd{=}\hlnum{3}\hlopt{/}\hlnum{0.5}\hlstd{,} \hlstr{"sigma.sq"}\hlstd{=}\hlnum{1}\hlstd{,} \hlstr{"tau.sq"}\hlstd{=}\hlnum{1}\hlstd{)}
\hlstd{R> }\hlstd{priors} \hlkwb{<-} \hlkwd{list}\hlstd{(}\hlstr{"phi.Unif"}\hlstd{=}\hlkwd{c}\hlstd{(}\hlnum{3}\hlopt{/}\hlnum{1}\hlstd{,} \hlnum{3}\hlopt{/}\hlnum{0.1}\hlstd{),} \hlstr{"sigma.sq.IG"}\hlstd{=}\hlkwd{c}\hlstd{(}\hlnum{2}\hlstd{,} \hlnum{1}\hlstd{),}
\hlstd{+ }  \hlstr{"tau.sq.IG"}\hlstd{=}\hlkwd{c}\hlstd{(}\hlnum{2}\hlstd{,} \hlnum{1}\hlstd{))}
\hlstd{R> }\hlstd{cov.model} \hlkwb{<-} \hlstr{"exponential"}
\hlstd{R> }\hlstd{tuning} \hlkwb{<-} \hlkwd{list}\hlstd{(}\hlstr{"phi"}\hlstd{=}\hlnum{0.2}\hlstd{)}
\hlstd{R> }\hlstd{ord} \hlkwb{<-} \hlkwd{order}\hlstd{(coords[,}\hlnum{1}\hlstd{]}\hlopt{+}\hlstd{coords[,}\hlnum{2}\hlstd{])}
\hlstd{R> }\hlstd{sim.s} \hlkwb{<-} \hlkwd{spNNGP}\hlstd{(}\hlkwc{formula}\hlstd{=y}\hlopt{~}\hlstd{x,} \hlkwc{coords}\hlstd{=coords,} \hlkwc{starting}\hlstd{=starting,}
\hlstd{+ }  \hlkwc{tuning}\hlstd{=tuning,} \hlkwc{priors}\hlstd{=priors,} \hlkwc{cov.model}\hlstd{=cov.model,} \hlkwc{n.samples}\hlstd{=n.samples,}
\hlstd{+ }  \hlkwc{n.neighbors}\hlstd{=}\hlnum{10}\hlstd{,} \hlkwc{method}\hlstd{=}\hlstr{"latent"}\hlstd{,} \hlkwc{ord}\hlstd{=ord,} \hlkwc{n.omp.threads}\hlstd{=}\hlnum{4}\hlstd{,}
\hlstd{+ }  \hlkwc{n.report}\hlstd{=}\hlnum{1000}\hlstd{,} \hlkwc{fit.rep}\hlstd{=}\hlnum{TRUE}\hlstd{,} \hlkwc{sub.sample}\hlstd{=}\hlkwd{list}\hlstd{(}\hlkwc{start}\hlstd{=}\hlnum{1000}\hlstd{),}
\hlstd{+ }  \hlkwc{return.neighbor.info} \hlstd{=} \hlnum{TRUE}\hlstd{)}
\end{alltt}
\begin{verbatim}
## ----------------------------------------
## 	Building the neighbor list
## ----------------------------------------
## ----------------------------------------
## Building the neighbors of neighbors list
## ----------------------------------------
## ----------------------------------------
## 	Model description
## ----------------------------------------
## NNGP Latent model fit with 5000 observations.
## 
## Number of covariates 2 (including intercept if specified).
## 
## Using the exponential spatial correlation model.
## 
## Using 10 nearest neighbors.
## 
## Number of MCMC samples 2000.
## 
## Priors and hyperpriors:
## 	beta flat.
## 	sigma.sq IG hyperpriors shape=2.00000 and scale=1.00000
## 	tau.sq IG hyperpriors shape=2.00000 and scale=1.00000
## 	phi Unif hyperpriors a=3.00000 and b=30.00000
## 
## Source compiled with OpenMP support and model fit using 4 thread(s).
## ----------------------------------------
## 		Sampling
## ----------------------------------------
## Sampled: 1000 of 2000, 50.00%
## Report interval Metrop. Acceptance rate: 46.50%
## Overall Metrop. Acceptance rate: 46.50%
## -------------------------------------------------
## Sampled: 2000 of 2000, 100.00%
## Report interval Metrop. Acceptance rate: 33.70%
## Overall Metrop. Acceptance rate: 40.10%
## -------------------------------------------------
\end{verbatim}
\end{kframe}
\end{knitrout}

As seen in the output above the call to \code{spNNGP} prints some basic model and sampler information. The output also notes \texttt{Source compiled with OpenMP support and model fit using 4 thread(s)}. \pkg{spNNGP} functions will throw a warning if \proglang{R} was not compiled with \proglang{openMP} support and \code{n.omp.threads} is set to a value greater than 1.
  
Objects returned by \code{spNNGP} and other functions in the package use S3 methods \code{summary} and \code{print}. As illustrated later, S3 methods for \code{fitted} and \code{residuals} are also implemented. The \code{print} method prints the initial call to the function as well as some model and sampler specifics. The default behavior of \code{summary} is to print posterior summaries for model parameters using samples from the second half of the MCMC chains. The values for arguments \code{sub.sample} and \code{quantiles} passed to \code{summary} allow for finer control on posterior summaries. Alternatively the user can access all posterior samples as \pkg{coda} \citep{coda} \code{mcmc} class objects in the \code{spNNGP} return objects' \code{p.beta.samples}, \code{p.theta.samples}, and \code{p.w.samples}, which as the names suggest hold $\bbeta$, $\btheta$, and $\bw$ samples, respectively.

\begin{knitrout}
\definecolor{shadecolor}{rgb}{0.969, 0.969, 0.969}\color{fgcolor}\begin{kframe}
\begin{alltt}
\hlstd{R> }\hlkwd{summary}\hlstd{(sim.s)}
\end{alltt}
\begin{verbatim}
## 
## Call:
## spNNGP(formula = y ~ x, coords = coords, method = "latent", 
##     n.neighbors = 10, starting = starting, tuning = tuning, 
##     priors = priors, cov.model = cov.model, n.samples = n.samples, 
##     n.omp.threads = 4, ord = ord, return.neighbor.info = TRUE, 
##     fit.rep = TRUE, sub.sample = list(start = 1000), n.report = 1000)
## 
## Model class is NNGP, method latent, family gaussian.
## 
## Model object contains 2000 MCMC samples.
## 
## Chain sub.sample:
## start = 1000
## end = 2000
## thin = 1
## samples size = 1001
##             2.5%    25%     50%     75%     97.5%  
## (Intercept) 0.5648  0.5948  0.6182  0.6757  0.7781 
## x           -0.1066 -0.0966 -0.0912 -0.0852 -0.0756
## sigma.sq    0.8535  0.9446  1.0013  1.1486  1.6633 
## tau.sq      0.2292  0.2378  0.2428  0.2486  0.2581 
## phi         3.5656  4.7293  5.8478  6.2808  6.9279
\end{verbatim}
\end{kframe}
\end{knitrout}

As requested by setting \code{fit.rep=TRUE} the \code{spNNGP} return object also holds samples for the regression fitted values, labeled \code{y.hat.samples}, and replicated data, labeled \code{y.rep.samples}. For convenience, the median and lower and upper 95\% credible intervals for MCMC samples at each location are provided in \code{y.hat.quants} and \code{y.rep.quants}. These samples and corresponding summaries can be accessed directly in the \code{spNNGP} return object or extracted using the S3 \code{fitted} method. Beyond simply extracting the regression fitted values and replicated data from \code{spNNGP} and other model objects in the package, the \code{fitted} function performs additional composition sampling if the requested MCMC sample subset differs from the one initially specified in the model call. For example, the initial call to \code{spNNGP} specified \code{sub.sample=list(start=1000)}, but if later we decide we want regression fitted values and replicated data for every 10$^{th}$ MCMC sample starting at sample 100, a call to \code{fitted(sim.s, sub.sample=list(start=100, thin=10))} would generate the desired subset (the \code{residuals} function provides the same behavior).

\subsubsection{MCMC-free inference}\label{sec:mcmc_free}

The conjugate model is called using the \code{spConjNNGP} function. Fixed $\alpha$, $\phi$, and perhaps $\nu$ are specified using a named vector passed to the \code{theta.alpha} argument. Alternatively, a $K$-fold cross-validation (where $K$ is set via the \code{k-fold} argument) is used to discover the ``optimal'' set of parameters if \code{theta.alpha} is passed a matrix with columns named \code{alpha}, \code{phi}, and perhaps \code{nu}. The ``optimal'' set of parameter values (i.e., a row in \code{theta.alpha}) is the one that minimizes the average value of the specified scoring rule over the $K$ folds. This scoring rule is set via the \code{score.rule} argument with options \code{"rmspe"} and \code{"crps"} for root mean squared prediction error (RMSPE) and continuous ranked probability score \cite[CRPS;][]{Gneiting07}. The $K$-fold cross-validation progress is printed to the screen as illustrated below. Once the optimal parameter set is identified, a final model is fit using all the available data. The description of this final model is given in the \code{Model description} section followed by the optimal set of $\alpha$, $\phi$, and, if the Mat{\'e}rn correlation model is used, $\nu$. 

The printout following the call to \code{spConjNNGP} below also includes a section called \code{Computing replicates} that reports on the exact sampling from the model parameters and regression fitted values posterior distributions, and generation of replicated data. Posterior sampling and generation of replicated data is optional and controlled by the \code{fit.rep} and \code{n.samples} arguments, with \code{n.samples} being set to the number of desired samples to collect. When \code{fit.rep=TRUE}, \code{spConjNNGP} effectively calls the S3 method \code{fitted} function for the conjugate model class. Hence, if posterior samples are not collected in the initial call to \code{spConjNNGP} or a different number of samples is needed, then a call to \code{fitted} using the \code{spConjNNGP} object will generate the required samples. 

\begin{knitrout}
\definecolor{shadecolor}{rgb}{0.969, 0.969, 0.969}\color{fgcolor}\begin{kframe}
\begin{alltt}
\hlstd{R> }\hlstd{theta.alpha} \hlkwb{<-} \hlkwd{as.matrix}\hlstd{(}\hlkwd{expand.grid}\hlstd{(}\hlkwd{seq}\hlstd{(}\hlnum{0.01}\hlstd{,}\hlnum{1}\hlstd{,}\hlkwc{length.out}\hlstd{=}\hlnum{15}\hlstd{),}
\hlstd{+ }  \hlkwd{seq}\hlstd{(}\hlnum{3}\hlstd{,}\hlnum{30}\hlstd{,}\hlkwc{length.out}\hlstd{=}\hlnum{15}\hlstd{)))}
\hlstd{R> }\hlkwd{colnames}\hlstd{(theta.alpha)} \hlkwb{<-} \hlkwd{c}\hlstd{(}\hlstr{"alpha"}\hlstd{,}\hlstr{"phi"}\hlstd{)}
\hlstd{R> }\hlstd{sim.c} \hlkwb{<-} \hlkwd{spConjNNGP}\hlstd{(y}\hlopt{~}\hlstd{x,} \hlkwc{coords}\hlstd{=}\hlkwd{as.matrix}\hlstd{(coords),}
\hlstd{+ }  \hlkwc{cov.model}\hlstd{=}\hlstr{"exponential"}\hlstd{,} \hlkwc{sigma.sq.IG}\hlstd{=}\hlkwd{c}\hlstd{(}\hlnum{2}\hlstd{,}\hlnum{0.5}\hlopt{*}\hlkwd{var}\hlstd{(y)),}
\hlstd{+ }  \hlkwc{n.neighbors}\hlstd{=}\hlnum{15}\hlstd{,} \hlkwc{ord}\hlstd{=ord,}
\hlstd{+ }  \hlkwc{theta.alpha}\hlstd{=theta.alpha,}
\hlstd{+ }  \hlkwc{k.fold} \hlstd{=} \hlnum{2}\hlstd{,} \hlkwc{score.rule} \hlstd{=} \hlstr{"rmspe"}\hlstd{,}
\hlstd{+ }  \hlkwc{fit.rep}\hlstd{=}\hlnum{TRUE}\hlstd{,} \hlkwc{n.samples}\hlstd{=}\hlnum{200}\hlstd{,}
\hlstd{+ }  \hlkwc{n.omp.threads}\hlstd{=}\hlnum{4}\hlstd{)}
\end{alltt}
\begin{verbatim}
## ----------------------------------------
## 	Starting k-fold
## ----------------------------------------
## 
  |                                    
  |                              |   0%
  |                                    
  |***************               |  50%
  |                                    
  |******************************| 100%
## ----------------------------------------
## 	Model description
## ----------------------------------------
## NNGP Conjugate model fit with 5000 observations.
## 
## Number of covariates 2 (including intercept if specified).
## 
## Using the exponential spatial correlation model.
## 
## Using 15 nearest neighbors.
## 
## Source compiled with OpenMP support and model fit using 4 thread(s).
## ------------
## Priors and hyperpriors:
## 	beta flat.
## 	sigma.sq IG hyperpriors shape=2.00000 and scale=0.56696
## ------------
## 	Estimation for parameter set(s)
## Set phi=4.92857 and alpha=0.22214
## ----------------------------------------
## 	Computing replicates
## ----------------------------------------
## NNGP Response model fit with 5000 observations.
## 
## Number of covariates 2 (including intercept if specified).
## 
## Using the exponential spatial correlation model.
## 
## Using 15 nearest neighbors.
## 
## Number of MCMC samples 200.
## 
## Source compiled with OpenMP support and model fit using 4 thread(s).
## ------------
## 		Sampling
## Sampled: 100 of 200, 50.00%
## Sampled: 200 of 200, 100.00%
\end{verbatim}
\end{kframe}
\end{knitrout}

Results from the $K$-fold cross-validation are returned by \code{spConjNNGP} and held in the \code{k.fold.scores} matrix which is a copy of \code{theta.alpha} with additional columns for $K$-fold RMSPE and CRPS as illustrated below.
\begin{knitrout}
\definecolor{shadecolor}{rgb}{0.969, 0.969, 0.969}\color{fgcolor}\begin{kframe}
\begin{alltt}
\hlstd{R> }\hlkwd{head}\hlstd{(sim.c}\hlopt{$}\hlstd{k.fold.scores,} \hlkwc{n}\hlstd{=}\hlnum{3}\hlstd{)}
\end{alltt}
\begin{verbatim}
##      phi      alpha     rmspe      crps
## [1,]   3 0.01000000 0.6262589 0.3567326
## [2,]   3 0.08071429 0.6019226 0.3399307
## [3,]   3 0.15142857 0.6005768 0.3387949
\end{verbatim}
\end{kframe}
\end{knitrout}

\begin{figure}[!htp]
  \centering
    \subfigure[]{\includegraphics[trim=0cm 0cm 0cm 0cm,clip,width=7cm]{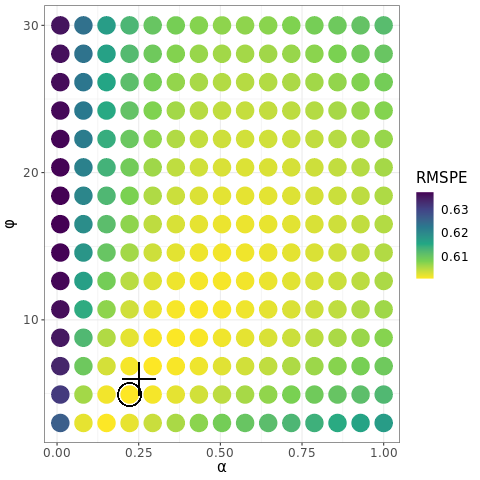}\label{simConjGridRMSPE}}
    \subfigure[]{\includegraphics[trim=0cm 0cm 0cm 0cm,clip,width=7cm]{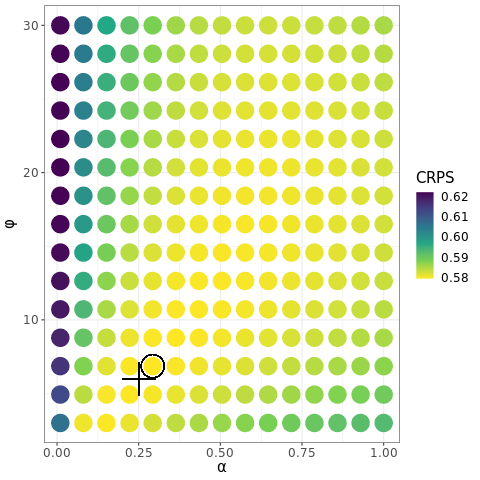}\label{simConjGridCRPS}}
  \caption{Simulated data analysis \code{spConjNNGP} parameter search grid and $K$-fold cross-validation results for RMSPE \subref{simConjGridRMSPE} and CRPS \subref{simConjGridCRPS} scoring rule. The ``optimal'' parameter
combination is circled. A plus symbol identifies the ``true'' $\alpha$ and $\phi$ used to
generate the data.}\label{simConjGrid}
\end{figure}

\code{k.fold.scores} is useful for assessing predictive performance sensitivity to choice of covariance parameters. Figure~\ref{simConjGridRMSPE} was created by plotting the search grid parameter values and their resulting minimum RMSPE. Note, the call to \code{spConjNNGP} specifies \code{score.rule = "rmspe"} which means the ``optimal'' $\alpha$, $\phi$, and, if \code{cov.model = "matern"}, $\nu$ will be the set that minimizes RMSPE. If one sets \code{score.rule = "crps"} the \code{k.fold.scores} can be used to identify the set of covariance parameters that minimize CRPS (the result of which is illustrated in Figure~\ref{simConjGridCRPS}).

The S3 \code{summary} method provides parameter point estimates using the optimal set and, if \code{n.samples} is specified, posterior summaries. The \code{sub.sample} argument can be used in \code{summary} if the \code{spConjNNGP} \code{fit.rep=FALSE} or if you wish for a posterior summary for a different number of \code{n.samples}. If \code{sub.sample} is specified in \code{summary}, the function returns a matrix of the requested number of samples. The output from \code{summary(sim.c)} which uses the initial number of 200 samples is given later in Table~\ref{simEstTable}.

\begin{table}[!ht]
  \begin{center}
    \caption{Simulated data analysis posterior summaries of median and 95\% credible interval from the three model types. }\label{simEstTable}
    \begin{tabular}{lccccc}
                      &      &  & \multicolumn{3}{c}{Model type} \\
                      & True  & Full GP &  \code{Latent} & \code{Response} & \code{Conjugate}\\
      \hline
      $\beta_0$  & 1 & 0.69 (0.24,  1.09) & 0.62 (0.56, 0.78) & 0.7 (0.27, 1.1)& 0.62 (0.12, 1.19)\\ 
      $\beta_x$  & -0.1 & -0.09 (-0.11,  -0.07)&-0.09 (-0.11, -0.08) & -0.09 (-0.11 , -0.08)& -0.09 (-0.11, -0.08)\\ 
      $\sigma^2$ & 1 & 0.78 (0.67,  0.98)&1 (0.85, 1.66) &  0.98 (0.72, 1.32)& 1.11 (1.06, 1.15)\\ 
      $\phi$     & 6 & 7.58 (5.93,  9.19)&5.85 (3.57, 6.93) & 5.95 (4.53, 8.16)& 4.93\\ 
      $\tau^2$   & 0.25 & 0.24 (0.23,  0.26)&0.24 (0.23, 0.26) & 0.24 (0.23, 0.26)& 0.25 (0.24, 0.26)\\ 
   \hline
    \end{tabular}
  \end{center}
\end{table}

\subsubsection{Model diagnostics and prediction}\label{sec:mod_diag_and_pred}

When passed a \code{spNNGP} or \code{spCongNNGP} object the \code{spDiag} function returns a list that, depending on the model \code{method}, includes some or all of the following elements:

\begin{itemize}
\item[\code{DIC}] a data frame holding the Deviance information criterion (DIC) and associated values defined by \cite{spiegelhalter2002bayesian}. The \code{DIC} data frame includes rows labeled \code{DIC} the criterion (lower is better), \code{D} a goodness of fit, and \code{pD} the effective number of parameters.
  
\item[\code{WAIC}] a data frame holding Watanabe-Akaike information criteria (WAIC) and associated values. The \code{WAIC} data frame includes rows labeled \code{LPPD} log pointwise predictive density, \code{P.1} penalty term defined in unnumbered equation above Equation (11) in \cite{gelman2013}, \code{P.2} an alternative penalty term defined in Equation (11), and the criteria \code{WAIC.1} and \code{WAIC.2} (lower is better) computed using \code{P.1} and \code{P.2}, respectively.
  
\item[\code{GPD}] a data frame holding the values needed to compute the predictive criterion $D = G+P$ defined by \cite{Gelfand98}. The \code{GPD} data frame includes rows labeled \code{G} a goodness of fit, \code{P} a penalty term, and \code{D} the criterion (lower is better).
  
\item[\code{GRS}] a scoring rule, see Equation 27 in \cite{Gneiting07} for details, with larger values of GRS indicating better model fit.
\end{itemize}

Among the four model comparison metrics, \code{DIC} and \code{WAIC} rely on the assumption that given all the parameters (including latent ones), the data points are conditionally independent. 
The response and conjugate NNGP models do not preserve this conditional independence structure and do not provide samples from the latent effect. Hence, it is not appropriate to compute WAIC and DIC for them. Comparisons across the models using \code{GPD} and \code{GRS} scores require generating replicate data. We have discussed in Section \ref{sec:rep} how replicates have fundamentally different interpretation for the latent and response models. For the former, the replicates are generated conditional on the latent random effects and hence are spatially correlated with the original data, whereas for the marginalized response model, the replicate is simply a new realization of a multivariate Gaussian distribution with the same mean and covariance structure as the original data. Hence, generally the \code{GPD} and \code{GRS} scores will be better for the latent models (as is evident in Table \ref{simDiagTable}). It is not advisable to compare the latent and response NNGP models using \code{GPD} and \code{GRS} as they represent different principles of replication. Finally, we can compare the response model with the conjugate model using \code{GPD} and \code{GRS} as they both use the same form of replicates. However, the conjugate model uses cross-validation to tune hyper-parameters, violating the principles of all these model comparison metrics tailored for classical Bayesian procedures, and it is difficult to interpret the model comparison values for it. 

\begin{figure}[!htp]
  \centering
    \includegraphics[trim=0cm 0cm 0cm 0cm,clip,width=12cm]{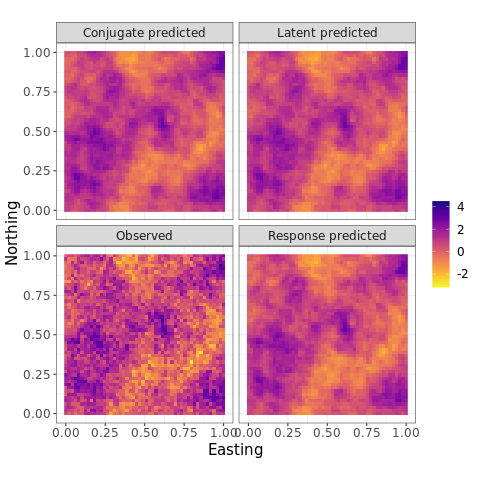}
  \caption{Observed and predicted simulated data $\by$ over the grid of holdout locations.}\label{simYSurfs}
\end{figure}

The code below calls \code{spDiag} for the \code{spNNGP} latent and \code{spConjNNGP} model output \code{sim.s} and \code{sim.c}, respectively. Additionally, for comparison, we ran \code{spNNGP} for the response model (i.e., by setting \code{method="response"}) and called the resulting object \code{sim.r}. Fit diagnostics for all three models are given in Table~\ref{simDiagTable}. However, as shown in Table~\ref{simEstTable}, all models recover the ``true'' parameter values well, and Figure~\ref{simYSurfs} shows all models produce comparable predictive surfaces.

\begin{knitrout}
\definecolor{shadecolor}{rgb}{0.969, 0.969, 0.969}\color{fgcolor}\begin{kframe}
\begin{alltt}
\hlstd{R> }\hlstd{s.diag} \hlkwb{<-} \hlkwd{spDiag}\hlstd{(sim.s)}
\hlstd{R> }\hlstd{r.diag} \hlkwb{<-} \hlkwd{spDiag}\hlstd{(sim.r)}
\hlstd{R> }\hlstd{c.diag} \hlkwb{<-} \hlkwd{spDiag}\hlstd{(sim.c)}
\end{alltt}
\end{kframe}
\end{knitrout}

\begin{table}[!ht]
  \begin{center}
    \caption{Simulated data analysis output from calls to \code{spDiag} for the three model types.}\label{simDiagTable}
    \begin{tabular}{lcccc}
                    & &\multicolumn{3}{c}{Model} \\
                    & Full GP &\code{Latent} & \code{Response} & \code{Conjugate}\\
      \hline
      \code{WAIC.1} & 8036.12 &8034.85 & -- & -- \\ 
      \code{WAIC.2} & 8501.09 &8499.76 & -- & -- \\ 
      \code{P.1}    & 915.32 &917.91 & -- & -- \\ 
      \code{P.2}    & 1147.8 &1150.36 & -- & -- \\ 
      \code{LPPD.2} & -3102.75 &-3099.52 & -- & -- \\     \hline
      \code{DIC}    & 8439.63 &8438.64 & -- & -- \\ 
      \code{pD}     & 1318.82 &1321.7 & -- & -- \\ 
      \code{L}      & -2900.99 &-2897.62 & -- & -- \\     \hline
      \code{G}      & 897.32 &894.85 & 5645.59 & 5912.89 \\
      \code{P}      & 1539.07 &1539.525 & 5234.422 & 6005.025 \\
      \code{D}      & 2436.39 &2434.377 & 10880.01 & 11917.92 \\     \hline
      \code{GRS}    & 2975.25 &2985.82 & -5662.55 & -5918.45 \\     \hline
    \end{tabular}
  \end{center}
\end{table}

\begin{figure}[!htp]
  \centering
    \includegraphics[trim=0cm 4cm 0cm 4cm,clip,width=12cm]{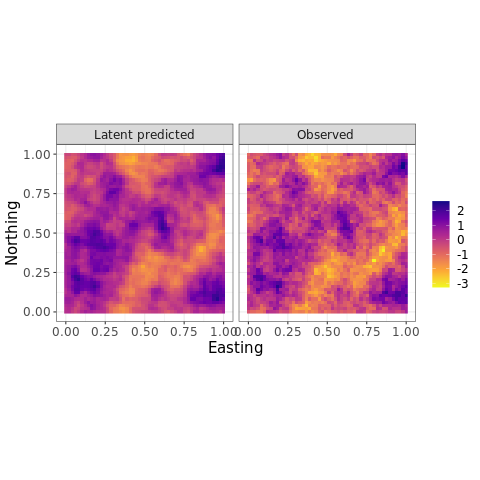}
  \caption{Observed and predicted simulated data $\bw$ over the grid of holdout locations.
  }\label{simWSurfs}
\end{figure}

Given the discordance in the interpretation of the traditional model comparison metrics for the NNGP models, a pragmatic way to compare them is based on their predictive performance on a hold out set. 
The \code{predict} function is used to generate posterior predictive samples for new locations with associated covariates, given a \code{spNNGP} or \code{spConjNNGP} object. The code below generates posterior predictive samples for the $n_0$=2500 holdout locations using the latent model. The latent model is the only \code{method} that provides posterior predictive samples for the latent effect $\bw$. These samples are held in the \code{p.w.0} matrix in the \code{s.pred} object generated below, and a summary of these samples along with the ``true'' $\bw$ are given in Figure~\ref{simWSurfs}. The \code{predict} function was also called for the response \code{sim.r} and conjugate \code{sim.c} model objects to generate posterior predictive samples for the holdout locations (held in output object \code{p.y.0}) along with subsequent surface summaries in Figure~\ref{simYSurfs}. 

\begin{knitrout}
\definecolor{shadecolor}{rgb}{0.969, 0.969, 0.969}\color{fgcolor}\begin{kframe}
\begin{alltt}
\hlstd{R> }\hlstd{s.pred} \hlkwb{<-} \hlkwd{predict}\hlstd{(sim.s,} \hlkwc{X.0}\hlstd{=}\hlkwd{cbind}\hlstd{(}\hlnum{1}\hlstd{,x.ho),} \hlkwc{coords.0}\hlstd{=coords.ho,}
\hlstd{+ }  \hlkwc{sub.sample}\hlstd{=}\hlkwd{list}\hlstd{(}\hlkwc{start}\hlstd{=}\hlnum{1000}\hlstd{,} \hlkwc{thin}\hlstd{=}\hlnum{10}\hlstd{),}
\hlstd{+ }  \hlkwc{n.omp.threads} \hlstd{=} \hlnum{4}\hlstd{,} \hlkwc{n.report}\hlstd{=}\hlnum{1000}\hlstd{)}
\end{alltt}
\begin{verbatim}
## ----------------------------------------
## 	Prediction description
## ----------------------------------------
## NNGP Latent model fit with 5000 observations.
## 
## Number of covariates 2 (including intercept if specified).
## 
## Using the exponential spatial correlation model.
## 
## Using 10 nearest neighbors.
## 
## Number of MCMC samples 101.
## 
## Predicting at 2500 locations.
## 
## 
## Source compiled with OpenMP support and model fit using 4 threads.
## -------------------------------------------------
## 		Predicting
## -------------------------------------------------
## Location: 1000 of 2500, 40.00%
## Location: 2000 of 2500, 80.00%
## Location: 2500 of 2500, 100.00%
\end{verbatim}
\end{kframe}
\end{knitrout}

\subsection{Timing given $n$ and number of cores}\label{sec:runtime}

\begin{figure}[!htp]
  \centering
  \subfigure{\includegraphics[trim=0cm 0cm 0cm 0cm,clip,width=7cm]{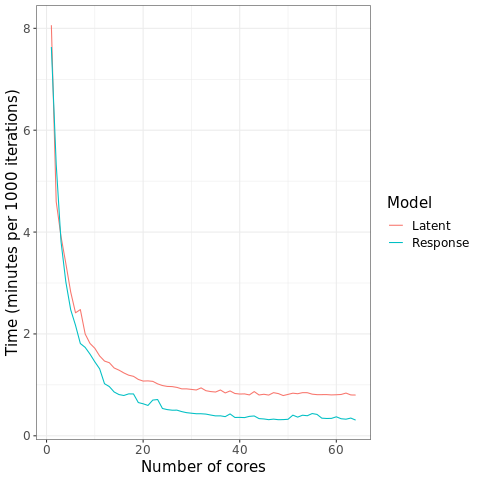}\label{cpuTiming}}
  \subfigure{\includegraphics[trim=0cm 0cm 0cm 0cm,clip,width=7cm]{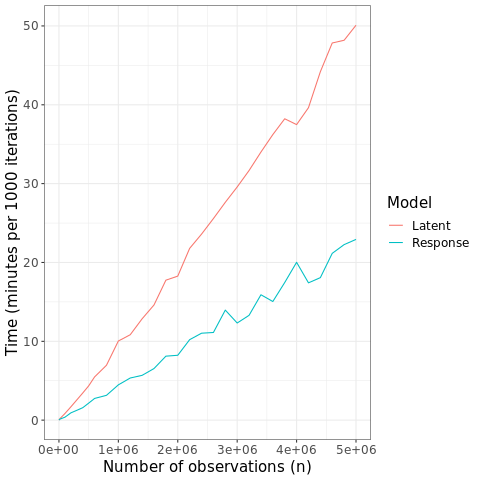}\label{nTiming}}
  \caption{\subref{cpuTiming} Runtime for 1000 MCMC iterations for $n$=100000 and differnt number of cores. \subref{nTiming} Runtime for 1000 MCMC iterations using 40 cores and $n$ from 1000 to 5 million.}\label{timing}
\end{figure}

Here we provide a brief overview of the relationship between $n$, model type, and number of cores. The computer used for these runtime experiments (and analysis in subsequent sections) is running a linux operating system with a AMD Ryzen Threadripper 3990X 64-Core Processor (128 threads) and \proglang{R} compiled with \proglang{openMP} with thread-enabled MKL 2019.5.281 build with \code{MKL_DEBUG_CPU_TYPE=5} \citep{intelMKL}. Timings generated by \code{proc.time()} is returned by core \pkg{spNNGP} functions in the \code{run.time} vector. Figure~\ref{cpuTiming} provides a sense of runtime (i.e., ``wall time'') needed to collect 1000 MCMC samples for $n$=100000 using the response and latent algorithms for a range of cores. Execution time for the conjugate model is about equal to one MCMC iteration of the response model. As this figure shows, for this computer and $n$, there is little speed-up beyond $\sim$40 cores mostly due to communication overhead. Then fixing the number of cores at 40, Figure~\ref{nTiming} gives a sense of computing time required for different size $n$.

\subsection{Analysis of forest canopy height}\label{sec:lidar-data}
\begin{figure}[!htp]
 \centering
 	\subfigure[]{\includegraphics[trim=0cm 3.75cm 0cm 3cm,clip,width=8.9cm]{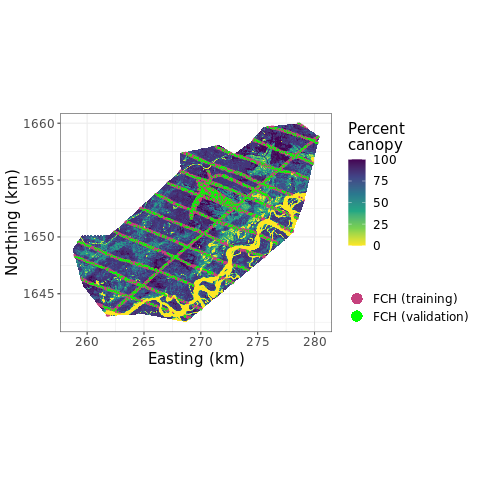}\label{bcef-map}}
        \subfigure[]{\includegraphics[trim=0cm 0cm 0cm 0cm,clip,width=6.5cm]{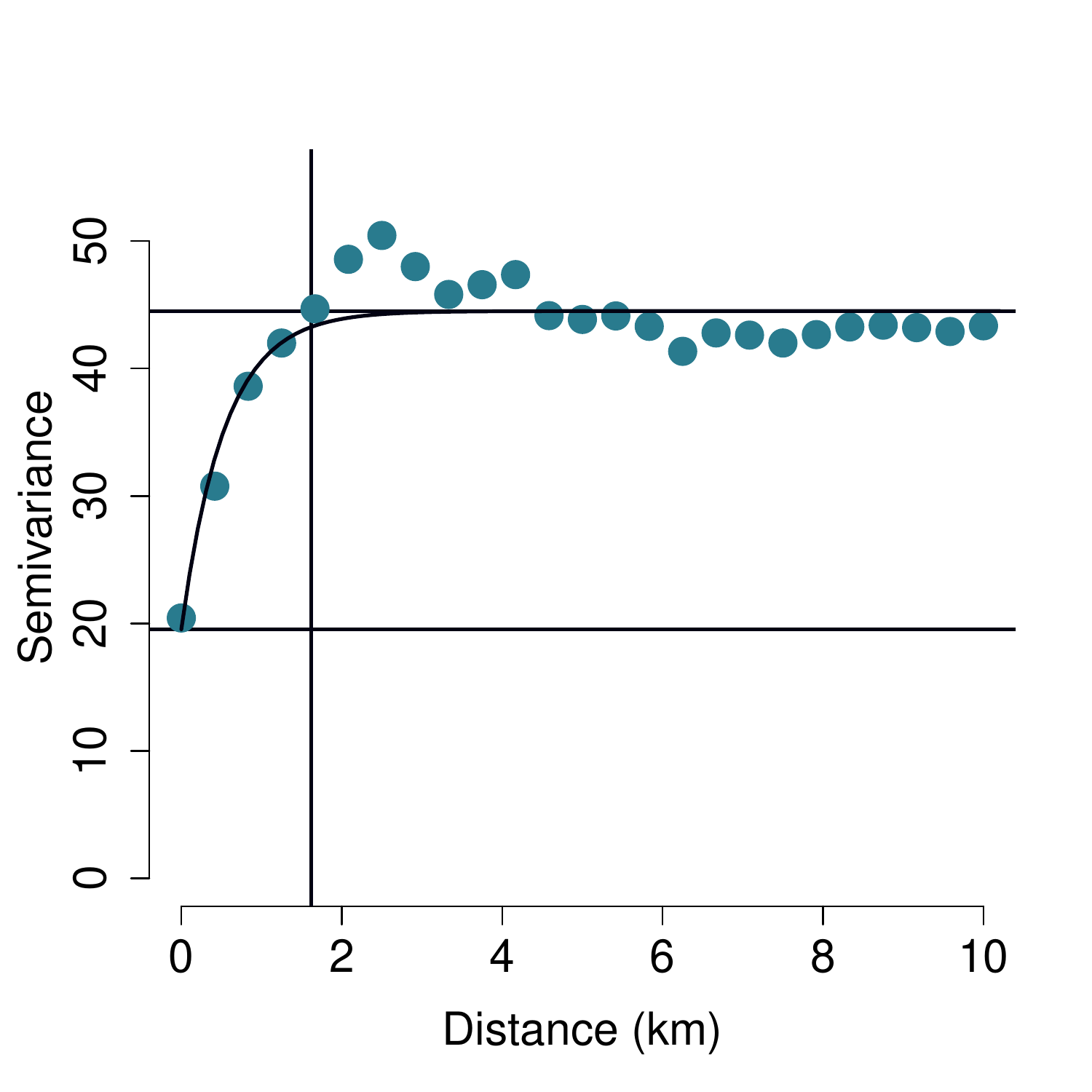}\label{bcef-vario}}
     	\caption{\subref{bcef-map} Bonanza Creek Experimental Forest (BCEF) with G-LiHT LiDAR forest canopy height (FCH) estimates for model training and validation. The underlying map is the percent tree cover (PTC). \subref{bcef-vario} Semivariogram of BCEF non-spatial regression model residuals. Exponential covariance function estimate denoted by the curved line with associated estimates for $\tau^2$, $\sigma^2$, and the effective spatial range are given by the lower horizontal, upper horizontal, and vertical lines, respectively.}\label{bcef}
\end{figure}

In this section we analyze a forest canopy height dataset at $n$=188,717 locations using \pkg{spNNGP}. Digital maps of forest structure are key inputs to many ecosystem and Earth system modeling efforts \citep{finney04, Hurtt04, stratton06, lefsky10, klein15}. These and similar applications seek inference about forest canopy height variables and predictions that can be propagated through computer models of ecosystem function to yield more robust error quantification. Given the scientific and applied interest in forest structure, there is increasing demand for wall-to-wall (i.e., complete domain coverage) forest canopy height data at national and biome scales. Next generation LiDAR systems capable of large-scale mapping of forest canopy characteristics, such as ICESat-2 \citep{abdalati2010, ICESAT2}, Global Ecosystem Dynamics Investigation LiDAR \citep{GEDI2014}, and NASA Goddard's LiDAR, Hyperspectral, and Thermal (G-LiHT) Airborne Imager \citep{cook2013}, sample forest features using LiDAR instruments in long transects or cluster designs (see, e.g., the strips of LiDAR in Figure~\ref{bcef-map}). These next generation systems yield LiDAR data over the desired large spatial extents; however, the sparseness of the LiDAR sampling designs means prediction is required to deliver the desired wall-to-wall data products.

Our goal is to create high spatial resolution forest canopy height predictions, with accompanying uncertainty estimates for the Bonanza Creek Experimental Forest (BCEF; \url{https://www.lter.uaf.edu}) located in interior Alaska, USA. The BCEF domain delineated for this study, Figure~\ref{bcef-map}, is $\sim$21,000 ha and includes a section of the Tanana River floodplain along the southeastern border. The BCEF is a mixture of non-forest and forest vegetation featuring white spruce, black spruce, tamarack, quaking aspen, and balsam poplar trees mixed with willow and alder shrubland species \cite{BCEF16}. Figure~\ref{bcef-map} also shows location of the $n$=188,717 G-LiHT LiDAR forest canopy height (FCH) estimates. These data are included in the \pkg{spNNGP} package. 

The 188,717 FCH estimates come from the G-LiHT LiDAR point cloud summarized to a $13\times 13$ m grid cell size \citep{GLIHT2016}. Over each grid cell, the maximum canopy height (i.e., FCH) was estimated using the 100$^{\text{th}}$ percentile height of the point cloud. A Landsat derived percent tree cover (PTC) data product developed by \cite{hansen13}, shown as the underlying surface in Figure~\ref{bcef-map} is used as a predictor variable for FCH. PTC is the percent tree cover estimates for peak growing season in 2010 and was created using a regression tree model applied to Landsat 7 ETM+ annual composites.

A semivariogram of the non-spatial regression model residuals can inform how the residual spatial/non-spatial variance (i.e., outcome variance not explained by the regression mean) is partitioned and the spatial range, see, e.g., Chapter 5 in \cite{BCG14} for details. Here we consider the residuals from
\begin{equation}\label{bcef-ns}
  \by = \beta_0 + \beta_{PTC}\bx + \bepsilon,
\end{equation}
where $\by$ is the vector of observed FCH estimates, $\beta_0$ is an intercept, $\beta_{PTC}$ is the slope coefficient associated with the PTC predictor variable denoted as $\bx$, and $\bepsilon$ is the $n\times 1$ vector following $N(\bzero, \tau^2\bI_n)$. In the subsequent analyses we use an exponential spatial correlation function that approaches zero as the distance between locations increases. Therefore we define the distance, $d_0$, at which this correlation drops to 0.05 as the ``effective spatial range,'' which allows us to solve $\phi=-\log(0.05)/d_0  \approx 3/d_0$ . Using the \code{variog} and \code{variofit} functions in the \pkg{geoR} package \citep{geoR}, the semivarogram and empirical parameter estimates for the BCEF are given in Figure~\ref{bcef-vario}. Due to computational constraints we used a random subset of 25,000 residuals from (\ref{bcef-ns}) to generate the variogram.

\subsubsection{Estimation and prediction}\label{sec:est_and_pred}

Here we consider the non-spatial, latent, response, and conjugate models for BCEF data. Posterior samples for the non-spatial regression model were generated using the \code{bayesLMRef} function in \pkg{spBayes}. Models are assessed using output from \code{spDiag}. Further, out-of-sample predictive performance was assessed by fitting the models to 100,000 observations (selected at random from the 188,717) and then predicting for the remaining holdout 88,717 observations. Finally, the models are used to predict \code{FCH} for a grid of 237,617 locations over the BCEF where \code{PTC} was recorded and resulting maps are compared.

\begin{knitrout}
\definecolor{shadecolor}{rgb}{0.969, 0.969, 0.969}\color{fgcolor}\begin{kframe}
\begin{alltt}
\hlstd{R> }\hlstd{n.samples} \hlkwb{<-} \hlnum{5000}
\hlstd{R> }\hlstd{starting} \hlkwb{<-} \hlkwd{list}\hlstd{(}\hlstr{"phi"}\hlstd{=}\hlnum{3}\hlopt{/}\hlnum{2}\hlstd{,} \hlstr{"sigma.sq"}\hlstd{=}\hlnum{40}\hlstd{,} \hlstr{"tau.sq"}\hlstd{=}\hlnum{1}\hlstd{)}
\hlstd{R> }\hlstd{priors} \hlkwb{<-} \hlkwd{list}\hlstd{(}\hlstr{"phi.Unif"}\hlstd{=}\hlkwd{c}\hlstd{(}\hlnum{3}\hlopt{/}\hlnum{10}\hlstd{,} \hlnum{3}\hlopt{/}\hlnum{0.1}\hlstd{),} \hlstr{"sigma.sq.IG"}\hlstd{=}\hlkwd{c}\hlstd{(}\hlnum{2}\hlstd{,} \hlnum{40}\hlstd{),}
\hlstd{+ }  \hlstr{"tau.sq.IG"}\hlstd{=}\hlkwd{c}\hlstd{(}\hlnum{2}\hlstd{,} \hlnum{10}\hlstd{))}
\hlstd{R> }\hlstd{cov.model} \hlkwb{<-} \hlstr{"exponential"}
\hlstd{R> }\hlstd{tuning} \hlkwb{<-} \hlkwd{list}\hlstd{(}\hlstr{"phi"}\hlstd{=}\hlnum{0.02}\hlstd{)}
\hlstd{R> }\hlstd{bcef.s} \hlkwb{<-} \hlkwd{spNNGP}\hlstd{(FCH}\hlopt{~}\hlstd{PTC,} \hlkwc{coords}\hlstd{=}\hlkwd{c}\hlstd{(}\hlstr{"x"}\hlstd{,}\hlstr{"y"}\hlstd{),} \hlkwc{data}\hlstd{=BCEF.mod,}
\hlstd{+ }  \hlkwc{starting}\hlstd{=starting,} \hlkwc{method}\hlstd{=}\hlstr{"latent"}\hlstd{,} \hlkwc{n.neighbors}\hlstd{=}\hlnum{10}\hlstd{,}
\hlstd{+ }  \hlkwc{tuning}\hlstd{=tuning,} \hlkwc{priors}\hlstd{=priors,} \hlkwc{cov.model}\hlstd{=cov.model,}
\hlstd{+ }  \hlkwc{n.samples}\hlstd{=n.samples,} \hlkwc{n.omp.threads}\hlstd{=}\hlnum{40}\hlstd{,} \hlkwc{n.report}\hlstd{=}\hlnum{2500}\hlstd{,}
\hlstd{+ }  \hlkwc{fit.rep}\hlstd{=}\hlnum{TRUE}\hlstd{,} \hlkwc{sub.sample}\hlstd{=}\hlkwd{list}\hlstd{(}\hlkwc{start}\hlstd{=}\hlnum{4000}\hlstd{,} \hlkwc{thin}\hlstd{=}\hlnum{10}\hlstd{))}
\end{alltt}
\begin{verbatim}
## ----------------------------------------
## 	Building the neighbor list
## ----------------------------------------
## ----------------------------------------
## Building the neighbors of neighbors list
## ----------------------------------------
## ----------------------------------------
## 	Model description
## ----------------------------------------
## NNGP Latent model fit with 100000 observations.
## 
## Number of covariates 2 (including intercept if specified).
## 
## Using the exponential spatial correlation model.
## 
## Using 10 nearest neighbors.
## 
## Number of MCMC samples 5000.
## 
## Priors and hyperpriors:
## 	beta flat.
## 	sigma.sq IG hyperpriors shape=2.00000 and scale=40.00000
## 	tau.sq IG hyperpriors shape=2.00000 and scale=10.00000
## 	phi Unif hyperpriors a=0.30000 and b=30.00000
## 
## Source compiled with OpenMP support and model fit using 40 thread(s).
## ----------------------------------------
## 		Sampling
## ----------------------------------------
## Sampled: 2500 of 5000, 50.00%
## Report interval Metrop. Acceptance rate: 35.12%
## Overall Metrop. Acceptance rate: 35.12%
## -------------------------------------------------
## Sampled: 5000 of 5000, 100.00%
## Report interval Metrop. Acceptance rate: 33.36%
## Overall Metrop. Acceptance rate: 34.24%
## -------------------------------------------------
\end{verbatim}
\end{kframe}
\end{knitrout}

For brevity, the output from subsequent calls to \code{spNNGP} for the response model and \code{spConjNNGP} are surpressed by setting \code{verbose=FALSE}. 

\begin{knitrout}
\definecolor{shadecolor}{rgb}{0.969, 0.969, 0.969}\color{fgcolor}\begin{kframe}
\begin{alltt}
\hlstd{R> }\hlstd{tuning} \hlkwb{<-} \hlkwd{list}\hlstd{(}\hlstr{"phi"}\hlstd{=}\hlnum{0.01}\hlstd{,} \hlstr{"sigma.sq"}\hlstd{=}\hlnum{0.01}\hlstd{,} \hlstr{"tau.sq"}\hlstd{=}\hlnum{0.005}\hlstd{)}
\hlstd{R> }\hlstd{bcef.r} \hlkwb{<-} \hlkwd{spNNGP}\hlstd{(FCH}\hlopt{~}\hlstd{PTC,} \hlkwc{coords}\hlstd{=}\hlkwd{c}\hlstd{(}\hlstr{"x"}\hlstd{,}\hlstr{"y"}\hlstd{),} \hlkwc{data}\hlstd{=BCEF.mod,}
\hlstd{+ } \hlkwc{starting}\hlstd{=starting,} \hlkwc{method}\hlstd{=}\hlstr{"response"}\hlstd{,} \hlkwc{n.neighbors}\hlstd{=}\hlnum{10}\hlstd{,}
\hlstd{+ }  \hlkwc{tuning}\hlstd{=tuning,} \hlkwc{priors}\hlstd{=priors,} \hlkwc{cov.model}\hlstd{=cov.model,}
\hlstd{+ }  \hlkwc{n.samples}\hlstd{=n.samples,} \hlkwc{n.omp.threads}\hlstd{=}\hlnum{40}\hlstd{,} \hlkwc{n.report}\hlstd{=}\hlnum{2500}\hlstd{,}
\hlstd{+ }  \hlkwc{fit.rep}\hlstd{=}\hlnum{TRUE}\hlstd{,} \hlkwc{sub.sample}\hlstd{=}\hlkwd{list}\hlstd{(}\hlkwc{start}\hlstd{=}\hlnum{4000}\hlstd{,} \hlkwc{thin}\hlstd{=}\hlnum{10}\hlstd{),}
\hlstd{+ }  \hlkwc{verbose}\hlstd{=}\hlnum{FALSE}\hlstd{)}
\end{alltt}
\end{kframe}
\end{knitrout}

\begin{knitrout}
\definecolor{shadecolor}{rgb}{0.969, 0.969, 0.969}\color{fgcolor}\begin{kframe}
\begin{alltt}
\hlstd{R> }\hlstd{theta.alpha} \hlkwb{<-} \hlkwd{as.matrix}\hlstd{(}\hlkwd{expand.grid}\hlstd{(}\hlkwd{seq}\hlstd{(}\hlnum{0.1}\hlstd{,}\hlnum{1}\hlstd{,}\hlkwc{length.out}\hlstd{=}\hlnum{15}\hlstd{),}
\hlstd{+ }  \hlkwd{seq}\hlstd{(}\hlnum{3}\hlopt{/}\hlnum{10}\hlstd{,}\hlnum{3}\hlopt{/}\hlnum{0.1}\hlstd{,}\hlkwc{length.out}\hlstd{=}\hlnum{15}\hlstd{)))}
\hlstd{R> }\hlkwd{colnames}\hlstd{(theta.alpha)} \hlkwb{<-} \hlkwd{c}\hlstd{(}\hlstr{"alpha"}\hlstd{,}\hlstr{"phi"}\hlstd{)}
\hlstd{R> }\hlstd{bcef.c} \hlkwb{<-} \hlkwd{spConjNNGP}\hlstd{(FCH}\hlopt{~}\hlstd{PTC,} \hlkwc{coords}\hlstd{=}\hlkwd{c}\hlstd{(}\hlstr{"x"}\hlstd{,}\hlstr{"y"}\hlstd{),} \hlkwc{data}\hlstd{=BCEF.mod,}
\hlstd{+ }  \hlkwc{cov.model}\hlstd{=}\hlstr{"exponential"}\hlstd{,} \hlkwc{sigma.sq.IG}\hlstd{=}\hlkwd{c}\hlstd{(}\hlnum{2}\hlstd{,} \hlnum{40}\hlstd{),}
\hlstd{+ }  \hlkwc{n.neighbors}\hlstd{=}\hlnum{10}\hlstd{,}
\hlstd{+ }  \hlkwc{theta.alpha}\hlstd{=theta.alpha,}
\hlstd{+ }  \hlkwc{k.fold} \hlstd{=} \hlnum{2}\hlstd{,} \hlkwc{score.rule} \hlstd{=} \hlstr{"crps"}\hlstd{,}
\hlstd{+ }  \hlkwc{fit.rep}\hlstd{=}\hlnum{TRUE}\hlstd{,} \hlkwc{n.samples}\hlstd{=}\hlnum{200}\hlstd{,}
\hlstd{+ }  \hlkwc{n.omp.threads}\hlstd{=}\hlnum{40}\hlstd{,}
\hlstd{+ }  \hlkwc{verbose}\hlstd{=}\hlnum{FALSE}\hlstd{)}
\end{alltt}
\end{kframe}
\end{knitrout}

\begin{figure}[!htp]
  \centering
    \includegraphics[trim=0cm 0cm 0cm 0cm,clip,width=7cm]{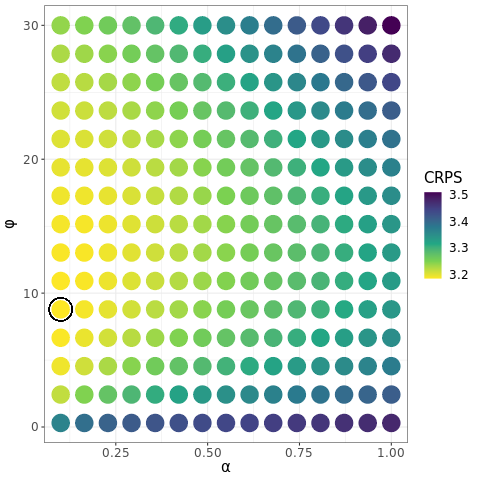}
  \caption{BCEF data analysis \code{spConjNNGP} parameter search grid and $K$-fold cross-validation results using CRPS scoring rulese. The ``optimal'' parameter
combination is circled.}\label{bcefConjGrid}
\end{figure}

Posterior summaries are given in Table~\ref{bcefEstTable}. As suggested by the exploratory variogram analysis, and now confirmed with formal model estimates, the spatial range is quite short, e.g., the latent model estimate of the median effective spatial range is $\sim$0.43 km (i.e., $-\log(0.05)/$6.91). Despite this short range, the spatial variance is large relative to the non-spatial variance. Such results are not surprising given the BCEF's composition and structure are the result of myriad large and small spatial scale biotic (e.g., insect disturbance) and abiotic (e.g., soil, topography, climate, wind, fire) factors that cause spatially complex mortality and regrowth patterns. Formal model fit diagnostics provided in Table~\ref{bcefDiagTable} suggests the addition of the latent spatial effect does improve fit compared with the non-spatial model.

\begin{table}[!ht]
  \begin{center}
    \caption{BCEF data analysis posterior summaries of median and 95\% credible interval from the three model types. }\label{bcefEstTable}
    \begin{tabular}{lcccc}
                      &        \multicolumn{4}{c}{Model} \\
                      & \code{Non-Spatial}   &  \code{Latent} & \code{Response} & \code{Conjugate}\\
      \hline
      $\beta_0$  & 1.36 (1.18, 1.49) &  11.09 (10.64, 11.92) & 9.94 (9.41, 10.47)& 10.45 (10.07, 10.85)\\ 
      $\beta_{PTC}$  & 0.2 (0.2, 0.2) & 0.03 (0.03, 0.03) & 0.03 (0.03 , 0.03)& 0.03 (0.03, 0.04)\\ 
      $\sigma^2$ & -- & 43.48 (41.71, 45.68) &  42.55 (40.6, 45.02)& 34.66 (34.33, 34.89)\\ 
      $\phi$     & -- & 6.91 (6.53, 7.22) & 7.19 (6.75, 7.52)& 4.93\\ 
      $\tau^2$   & 43.39 (43.12, 43.72) & 3.46 (3.35, 3.58) & 3.37 (3.27, 3.47)& 3.47 (3.43, 3.49)\\ 
   \hline
    \end{tabular}
  \end{center}
\end{table}

\begin{table}[!ht]
  \begin{center}
    \caption{BCEF data analysis model fit via \code{spDiag} and out-of-sampled prediction diagnostics for the non-spatial and three spatial model types fit to the BCEF data. The last four rows were calculated using prediction for the holdout set. The row labeled CI Cover is the percent of 95\% posterior predictive distribution credible intervals that cover the observed holdout value. The row labeled CI Width is the average width of the 95\% posterior predictive credible interval.}\label{bcefDiagTable}
    \begin{tabular}{lcccc}
                    & \multicolumn{4}{c}{Model type} \\
                    & \code{Non-Spatial} & \code{Latent} & \code{Response} & \code{Conjugate}\\
      \hline
      \code{WAIC.1} &660845 & 435074.6 & -- & -- \\ 
      \code{WAIC.2} &660845 & 462473.1 & -- & -- \\ 
      \code{P.1}    &2.35 & 27501.35 & -- & -- \\ 
      \code{P.2}    &2.38 & 41200.63 & -- & -- \\ 
      \code{LPPD.2} &-330420.1 & -190035.9 & -- & -- \\     \hline
      \code{DIC}    &660845.3 & 464395.5 & -- & -- \\ 
      \code{pD}     &2.69 & 56822.25 & -- & -- \\ 
      \code{L}      &-330420 & -175375.5 & -- & -- \\     \hline
      \code{G}      &4387671 & 152329.7 & 7086265 & 6403563 \\
      \code{P}      &4336921 & 542156.8 & 4119386 & 3502314 \\
      \code{D}      &8724591 & 694486.5 & 11205651 & 9905877 \\     \hline
      \code{GRS}    &-479230.7 & -196616.3 & -545860.2 & -539539.6 \\     \hline
      CRPS    &3.79 &1.57  & 1.9 &1.53  \\
      RMSPE    &6.62 &3.01  & 3.39 &2.94  \\
      CI Cover    &94.42 &85.8  & 72.56 &--  \\
      CI Width    &24.73 &7.82  & 5.76 &--  \\     \hline      
    \end{tabular}
  \end{center}
\end{table}

\begin{figure}[!htp]
  \centering
  \subfigure{\includegraphics[trim=0cm 0cm 0cm 0cm,clip,width=10cm]{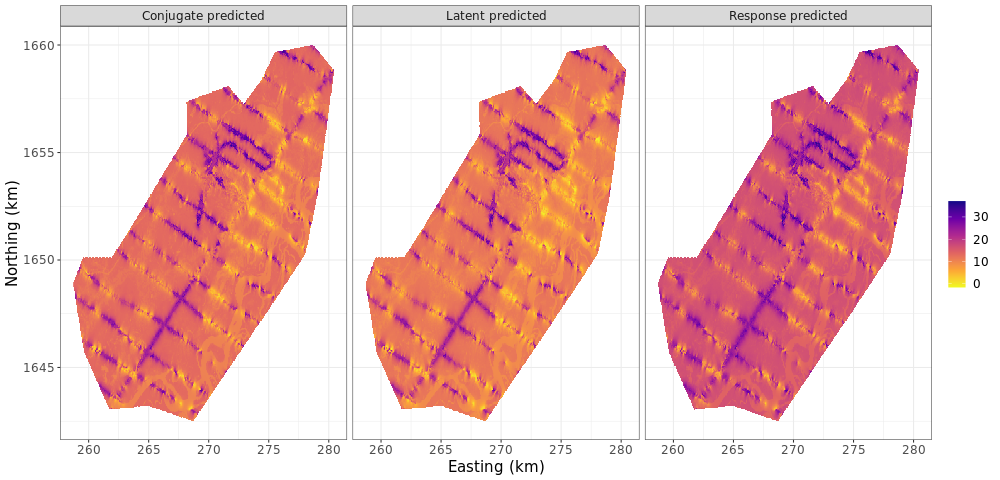}\label{bcefPredMu}}\\
  \subfigure{\includegraphics[trim=0cm 0cm 0cm 0cm,clip,width=10cm]{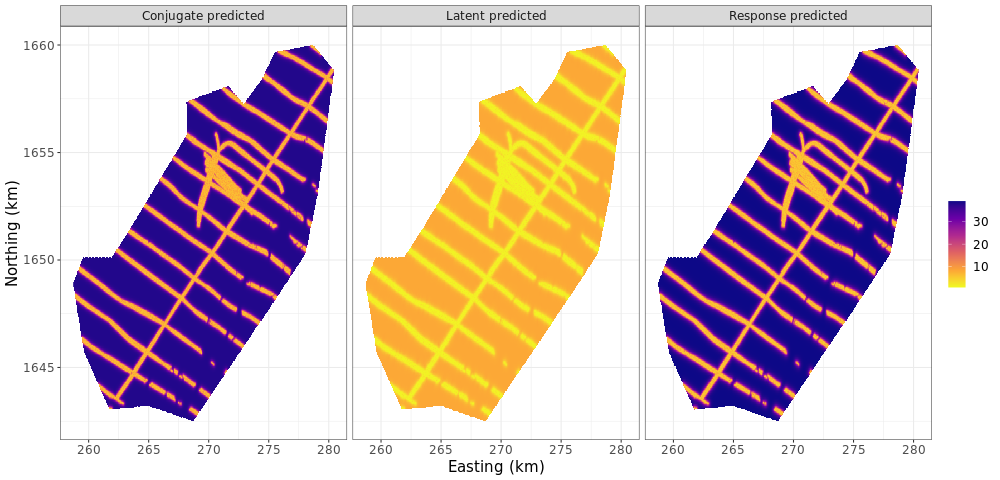}\label{bcefPredVar}}
  \caption{BCEF data analysis posterior predictive distribution mean \subref{bcefPredMu} and variance \subref{bcefPredVar}.}\label{bcefPred}
\end{figure}

Lastly, and as illustrated in the synthetic data analysis, a call to \code{predict} yields posterior predictive samples for the entire domain of interest which in this case is sampling over the grid of $n_0$=237617 locations where only PTC was recorded. A surface of the posterior distributions' mean and variance are given in Figures~\ref{bcefPredMu} and \ref{bcefPredVar}, respectively. Locations where observations are available to inform prediction (i.e., along flight lines) are clearly delineated by high prediction precision in the prediction variance surfaces Figure~\ref{bcefPredVar}. Given the relatively short spatial range this information borrowing to inform prediction does not extend too far off of the flight lines. These surfaces along with the out-of-sampled prediction performance metrics given in the last four rows in Table~\ref{bcefDiagTable} suggest there is not too much difference among the spatial models.

\subsection{Analysis of species distributions}\label{sec:spp-data}

In this Section, we present analysis of species distribution using the Binomial NNGP model. Species distribution models (SDMs) project the outcome of community assembly processes dispersal, the abiotic environment, and biotic factors onto geographic space \citep{Guisan2000,Pulliam2000}. Here, we reanalyze data recently presented in \cite{Lany2019} to develop a SDM for eastern hemlock (\emph{Tsuga canadensis L.}) coded as TSCA in subsequent analysis. The data comprise hemlock occurrence (Binomial outcome) on 17,743 forest stands across Michigan, USA. A set of predictors were also observed at each stand and subsequently used to explain the probability of hemlock occurrence. Predictor variables included minimum winter temperature (MIN), maximum summer temperature (MAX), total precipitation in the coldest quarter of the year (WIP), total precipitation in the warmest quarter of the year (SUP), annual actual evapotranspiration (AET) and annual climatic water deficit (DEF). 

We consider three candidate models: 1) non-spatial logistic regression using the P{\'o}lya-Gamma data-augmented sampler of \cite{polson2013bayesian} as implemented in the our \pkg{spNNGP} \code{PGLogit} function; 2) logistic regression with a space-varying Gaussian Predictive Process (GPP) random effect \cite{BGFS2008} using the \pkg{spBayes}' \code{spGLM} function; 3) logistic regression with space-varying NNGP again via the P{\'o}lya-Gamma sampler invoked using the \code{family="binomial"} argument in the \code{spNNGP} function as illustrated in the code below. Computational benefits of GPP models arise from the estimation of the latent Gaussian process at set of locations referred to as knots, where the number of knots is typically much smaller than $n$. When applying a GPP the analysis must strike an acceptable balance between computing time, which is governed by the number of knots, and Gaussian Process being approximated. More details about implementing a GPP can be found in \cite{BGFS2008}, \cite{finley2009}, \cite{Rajarshi11}, and \cite{Stein2014}. Here, for simplicity, we selected 25 knots on a fixed grid over the domain as it yielded a reasonable approximation to the desired GP with a computing time close that that of the NNGP.  All three models were fit using $n$=15,000 observations and assessed using goodness of fit metrics and out-of-sampled predictive performance based on a holdout set of $n_0$=2,743. 

Like a \code{spNNGP} class object, the return object from the \code{PGLogit} function can be passed to \code{spDiag} to yield model diagnostics as illustrated in the code below. Both \code{PGLogit} and \code{spNNGP} with \code{faimly="binomial"} accept different number of trials for each location, i.e., $n_i$ in Section~\ref{sec:pg}, which is passed via the \code{weights} argument; however, for the current setting we accept the default of all weights equal to 1, i.e., each stand yields either presence or absence of hemlock.

\begin{knitrout}
\definecolor{shadecolor}{rgb}{0.969, 0.969, 0.969}\color{fgcolor}\begin{kframe}
\begin{alltt}
\hlstd{R> }\hlstd{n.samples} \hlkwb{<-} \hlnum{10000}
\hlstd{R> }\hlstd{starting} \hlkwb{<-} \hlkwd{list}\hlstd{(}\hlstr{"phi"}\hlstd{=}\hlnum{3}\hlopt{/}\hlnum{50}\hlstd{,} \hlstr{"sigma.sq"}\hlstd{=}\hlnum{10}\hlstd{)}
\hlstd{R> }\hlstd{tuning} \hlkwb{<-} \hlkwd{list}\hlstd{(}\hlstr{"phi"}\hlstd{=}\hlnum{0.05}\hlstd{)}
\hlstd{R> }\hlstd{priors} \hlkwb{<-} \hlkwd{list}\hlstd{(}\hlstr{"phi.Unif"}\hlstd{=}\hlkwd{c}\hlstd{(}\hlnum{3}\hlopt{/}\hlnum{300}\hlstd{,} \hlnum{3}\hlopt{/}\hlnum{10}\hlstd{),} \hlstr{"sigma.sq.IG"}\hlstd{=}\hlkwd{c}\hlstd{(}\hlnum{2}\hlstd{,} \hlnum{10}\hlstd{))}
\hlstd{R> }\hlstd{cov.model} \hlkwb{<-} \hlstr{"exponential"}
\hlstd{R> }\hlstd{m.s} \hlkwb{<-} \hlkwd{spNNGP}\hlstd{(TSCA}\hlopt{~}\hlstd{MIN}\hlopt{+}\hlstd{MAX}\hlopt{+}\hlstd{SUP}\hlopt{+}\hlstd{WIP}\hlopt{+}\hlstd{AET}\hlopt{+}\hlstd{DEF,} \hlkwc{coords}\hlstd{=}\hlkwd{c}\hlstd{(}\hlstr{"long"}\hlstd{,}\hlstr{"lat"}\hlstd{),}
\hlstd{+ }  \hlkwc{data}\hlstd{=mi.mod,} \hlkwc{family}\hlstd{=}\hlstr{"binomial"}\hlstd{,} \hlkwc{method}\hlstd{=}\hlstr{"latent"}\hlstd{,} \hlkwc{n.neighbors}\hlstd{=}\hlnum{10}\hlstd{,}
\hlstd{+ }  \hlkwc{starting}\hlstd{=starting,} \hlkwc{tuning}\hlstd{=tuning,} \hlkwc{priors}\hlstd{=priors,}
\hlstd{+ }  \hlkwc{cov.model}\hlstd{=cov.model,} \hlkwc{n.samples}\hlstd{=n.samples,} \hlkwc{n.report}\hlstd{=}\hlnum{5000}\hlstd{,}
\hlstd{+ }  \hlkwc{fit.rep}\hlstd{=}\hlnum{TRUE}\hlstd{,} \hlkwc{sub.sample}\hlstd{=}\hlkwd{list}\hlstd{(}\hlkwc{start}\hlstd{=}\hlnum{9000}\hlstd{),} \hlkwc{n.omp.threads}\hlstd{=}\hlnum{40}\hlstd{)}
\end{alltt}
\begin{verbatim}
## ----------------------------------------
## 	Building the neighbor list
## ----------------------------------------
## ----------------------------------------
## Building the neighbors of neighbors list
## ----------------------------------------
## ----------------------------------------
## 	Model description
## ----------------------------------------
## NNGP Latent model fit with 15000 observations.
## 
## Number of covariates 7 (including intercept if specified).
## 
## Using the exponential spatial correlation model.
## 
## Using 10 nearest neighbors.
## 
## Number of MCMC samples 10000.
## 
## Priors and hyperpriors:
## 	beta flat.
## 	sigma.sq IG hyperpriors shape=2.00000 and scale=10.00000
## 	phi Unif hyperpriors a=0.01000 and b=0.30000
## 
## Source compiled with OpenMP support and model fit using 40 thread(s).
## ----------------------------------------
## 		Sampling
## ----------------------------------------
## Sampled: 5000 of 10000, 50.00%
## Report interval Metrop. Acceptance rate: 39.96%
## Overall Metrop. Acceptance rate: 39.96%
## -------------------------------------------------
## Sampled: 10000 of 10000, 100.00%
## Report interval Metrop. Acceptance rate: 39.78%
## Overall Metrop. Acceptance rate: 39.87%
## -------------------------------------------------
\end{verbatim}
\end{kframe}
\end{knitrout}

\begin{figure}[!htp]
  \centering
    \includegraphics[trim=0cm 0cm 0cm 0cm,clip,width=12cm]{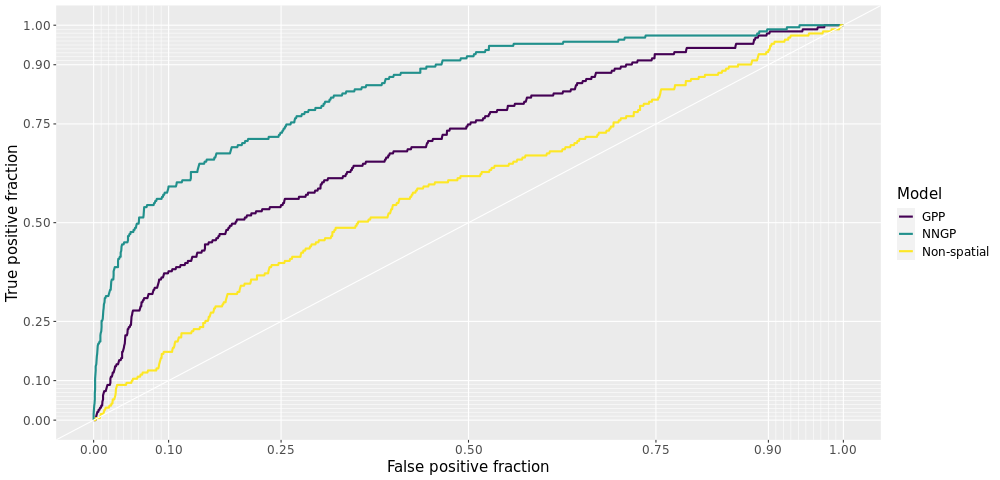}
  \caption{Michigan Eastern Hemlock SDM analysis out-of-sample posterior predictive mean ROC curves for the candidate models. Models with better prediction have curves closer to the top left corner.}\label{miROC}
\end{figure}

Parameter estimates for the three models are given in Table~\ref{miEstTable}, and shows that a number of the predictor variables help explain the probability of hemlock occurrence across the study area. Due to possible spatial confounding \citep{Hanks2015}, we should interpret the sign and significance of these regression parameters in the spatial models with caution. Within sample fit diagnostics given in Table~\ref{miDiagTable} and out-of-sample receiver operating characteristic (ROC) prediction curves (based on $n_0$=2,743 holdout locations) given in Figure~\ref{miROC}, suggest the latent spatial variables improve the species distribution models over that of the non-spatial model. Among the spatial models, the NNGP model outperforms the reduced rank predictive process model. The computing times per 1000 samples for the non-spatial, GPP, and NNGP models are 3, 13, and 9 seconds, respectively, using 10 cores.

\begin{table}[!ht]
  \begin{center}
    \caption{Michigan Eastern Hemlock SDM analysis posterior summaries of median and 95\% credible interval from the three candidate models.}\label{miEstTable}
    \begin{tabular}{lccc}
                      & \multicolumn{3}{c}{Model} \\
                      & \code{Non-Spatial}   &  \code{GPP 25 knots} & \code{NNGP Latent}\\
      \hline
   $\beta_0$   & -2.65 (-2.72, -2.59) & -3.6 (-3.71, -3.45) & -4.77 (-5.08, -4.47)\\ 
$\beta_{MIN}$  & 0.51 (0.39, 0.64) & 0.74 (0.56, 0.9) & 0.13 (-0.22, 0.42)\\ 
$\beta_{MAX}$  & -0.21 (-0.3, -0.12) & -0.12 (-0.26, 0.01) & -0.02 (-0.36, 0.21)\\ 
$\beta_{SUP}$  & 0.19 (0.11, 0.27) & -0.12 (-0.23, -0.03) & -0.16 (-0.44, 0.09)\\ 
$\beta_{WIP}$  & -0.01 (-0.1, 0.07) & 0.12 (-0.01, 0.25) & 0.01 (-0.27, 0.32)\\ 
$\beta_{AET}$  & -0.41 (-0.54, -0.27) & -0.32 (-0.49, -0.16) & -0.31 (-0.52, -0.1)\\ 
$\beta_{DEF}$  & -0.44 (-0.53, -0.34) & -0.35 (-0.48, -0.23) & -0.28 (-0.42, -0.13)\\ 
$\sigma^2$     & --                                                                                                                            & 4.12 (2.33, 8.46) & 8.45 (6.95, 9.79)\\ 
$\phi$         & --                                                                                                                            & 0.01 (0.01, 0.01) & 0.08 (0.06, 0.1)\\ 
   \hline
    \end{tabular}
  \end{center}
\end{table}

\begin{table}[!ht]
  \begin{center}
    \caption{Michigan Eastern Hemlock SDM analysis model fit via \code{spDiag} and out-of-sample prediction diagnostics for the non-spatial and two spatial models. }\label{miDiagTable}
    \begin{tabular}{lccc}
                    & \multicolumn{3}{c}{Model} \\
                    & \code{Non-Spatial} & \code{GPP 25 knots} & \code{NNGP Latent} \\
      \hline
      \code{WAIC.1} &7494.56 & 6889.75 &5252.05\\ 
      \code{WAIC.2} &7494.6 & 6890.07 & 5511.65\\ 
      \code{P.1}    &6.8 & 27.92 & 623.02\\ 
      \code{P.2}    &6.82 & 28.07 & 752.82\\ 
      \code{LPPD.2} &-3740.48 & -3416.96 & -2003 \\     \hline
      \code{DIC}    &7494.58 & 6888.86 & 5410.72\\ 
      \code{pD}     &6.82 & 27.02 & 781.7 \\   
    \end{tabular}
  \end{center}
\end{table}

\subsection{Statistical gap-filling of a massive remotely sensed data}\label{sec:ndvi-data}

\begin{figure}[!htp]
  \centering
  \subfigure[]{\includegraphics[trim=0cm 3.5cm 0cm 3cm,clip,width=7cm]{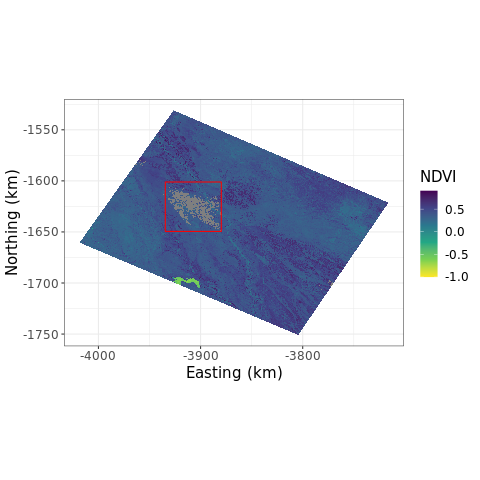}\label{kruger-ndvi}}
  \subfigure[]{\includegraphics[trim=0cm 3.5cm 0cm 3cm,clip,width=7cm]{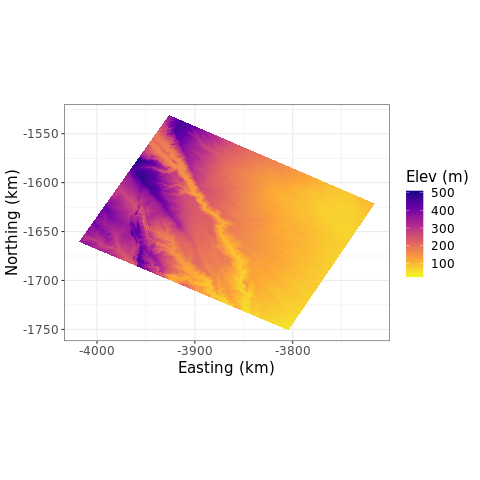}\label{kruger-elev}}\\
  \caption{\subref{kruger-ndvi} Normalized Difference Vegetation Index (NDVI) LandSat 8 sensor image over Limpopo National Park, Mozambique, Africa on 7/17/2015, with gray pixels indicate missing NDVI data. \subref{kruger-elev} land surface elevation used to help explain variation in NDVI.}\label{krugerData}
\end{figure}

As reviewed in Section~\ref{sec:intro}, methods and software are now emerging that can readily fit geostatistical models to data sets comprising locations in the 10s to 100s of thousands. However, as the number of observations climb into the millions, inferential and software options are quite limited. It is common to encounter data sets of such size in a variety of fields. For example, remotely sensed data of this magnitude either as gridded or scattered data products is now ubiquitous. Here we consider a massive-scale ``gap-filling'' exercise of missing Normalized Difference Vegetation Index (NDVI)---a measure of vegetation greenness--- data from around 39 million locations. The NDVI data for a LandSat 8 sensor image was taken over Limpopo National Park, Mozambique, Africa on 7/17/2015. The image shown in Figure~\ref{kruger-ndvi} has $n_0=$778644 pixels that are missing NDVI (denoted in gray) due to cloud cover during image acquisition. The red box in Figure~\ref{kruger-ndvi} delineates a region with a large number of missing pixels. Filling in missing NDVI values for this and other images in a time series over Limpopo was a step in a larger study conducted by \cite{Desanker2019} that looked at environmental drivers in vegetation phonology change. Our aim is to build a geostatistical model to predict the $n_0$ missing NDVI pixels given the $n$=38825052 observed NDVI pixels and complete coverage land surface elevation predictor variable shown in Figure~\ref{kruger-elev}.

While a MCMC solution is feasible via a call to \code{spNNGP} the runtime would be on the order of days, even on a multiprocessor computer. Therefore we opt for a MCMC-free approach using \code{spConjNNGP}---trading some inference about the spatial process parameters for substantially decreased computing time. An initial variogram analysis using a sample of residuals from NDVI regressed on elevation helped define \code{spConjNNGP}'s \code{theta.alpha} search grid of 100 $\phi$ and $\alpha$ combinations, and for setting the prior for $\sigma^2$. As noted in the code below, we used minimum CRPS to select the optimal set of covariance parameters (i.e., specified via the \code{score.rule} argument). 

\begin{knitrout}
\definecolor{shadecolor}{rgb}{0.969, 0.969, 0.969}\color{fgcolor}\begin{kframe}
\begin{alltt}
\hlstd{R> }\hlstd{theta.alpha} \hlkwb{<-} \hlkwd{as.matrix}\hlstd{(}\hlkwd{expand.grid}\hlstd{(}\hlkwd{seq}\hlstd{(}\hlnum{1.5}\hlstd{,} \hlnum{0.01}\hlstd{,} \hlkwc{length.out}\hlstd{=}\hlnum{10}\hlstd{),}
\hlstd{+ }  \hlkwd{seq}\hlstd{(}\hlnum{3}\hlopt{/}\hlnum{200}\hlstd{,} \hlnum{3}\hlopt{/}\hlnum{25}\hlstd{,} \hlkwc{length.out}\hlstd{=}\hlnum{10}\hlstd{)))}
\hlstd{R> }\hlkwd{colnames}\hlstd{(theta.alpha)} \hlkwb{<-} \hlkwd{c}\hlstd{(}\hlstr{"alpha"}\hlstd{,}\hlstr{"phi"}\hlstd{)}
\hlstd{R> }\hlstd{ser.c} \hlkwb{<-} \hlkwd{spConjNNGP}\hlstd{(ndvi}\hlopt{~}\hlstd{elev,} \hlkwc{coords}\hlstd{=}\hlkwd{c}\hlstd{(}\hlstr{"x"}\hlstd{,}\hlstr{"y"}\hlstd{),} \hlkwc{data}\hlstd{=ser.mod,}
\hlstd{+ }  \hlkwc{cov.model}\hlstd{=}\hlstr{"exponential"}\hlstd{,} \hlkwc{sigma.sq.IG}\hlstd{=}\hlkwd{c}\hlstd{(}\hlnum{2}\hlstd{,}\hlnum{0.01}\hlstd{),}
\hlstd{+ }  \hlkwc{n.neighbors}\hlstd{=}\hlnum{10}\hlstd{,}
\hlstd{+ }  \hlkwc{theta.alpha}\hlstd{=theta.alpha,}
\hlstd{+ }  \hlkwc{k.fold} \hlstd{=} \hlnum{2}\hlstd{,} \hlkwc{score.rule} \hlstd{=} \hlstr{"crps"}\hlstd{,}
\hlstd{+ }  \hlkwc{X.0} \hlstd{=} \hlkwd{cbind}\hlstd{(}\hlnum{1}\hlstd{,ser.ho}\hlopt{$}\hlstd{elev),}
\hlstd{+ }  \hlkwc{coords.0} \hlstd{=} \hlkwd{as.matrix}\hlstd{(ser.ho[,}\hlkwd{c}\hlstd{(}\hlstr{"x"}\hlstd{,}\hlstr{"y"}\hlstd{)]),}
\hlstd{+ }  \hlkwc{n.omp.threads}\hlstd{=}\hlnum{40}\hlstd{)}
\end{alltt}
\begin{verbatim}
## ----------------------------------------
## 	Starting k-fold
## ----------------------------------------
## 
  |                                    
  |                              |   0%
  |                                    
  |***************               |  50%
  |                                    
  |******************************| 100%
## ----------------------------------------
## 	Model description
## ----------------------------------------
## NNGP Conjugate model fit with 38825052 observations.
## 
## Number of covariates 2 (including intercept if specified).
## 
## Using the exponential spatial correlation model.
## 
## Using 10 nearest neighbors.
## 
## Source compiled with OpenMP support and model fit using 40 thread(s).
## ------------
## Priors and hyperpriors:
## 	beta flat.
## 	sigma.sq IG hyperpriors shape=2.00000 and scale=0.01000
## ------------
## Predicting at 778644 locations.
## ------------
## 	Estimation for parameter set(s)
## Set phi=0.12000 and alpha=0.01000
\end{verbatim}
\end{kframe}
\end{knitrout}

Given the size of this data set, the search time to identify nearest neighbor sets can take a considerable amount of time, even when using the \cite{Ra1993} fast code book algorithm invoked by \code{search.type="cb"}. In this case the search time was 8.86 minutes (search \code{proc.time()} is held in \code{neighbor.info$nn.indx.run.time} in the model object). Once the optimal $\phi$ and $\alpha$ was found, the runtime for parameter estimation and prediction for the missing pixels was 31.13 minutes. We have experimented with a variety of tree-based data structures (e.g., $kd$-trees modified to accommodate the nearest neighbor search constraints) and associated search algorithms, and believe there are substantial gains in search time efficiency to be had with additional development.

As shown below, a subsequent call to \code{summary} returns the optimal parameter set as well as point estimates for the other model parameters. Given the large size of this data set, parameter variance estimates are remarkably small (for some parameters the variance might be smaller than machine precision). The message at the end of the \code{summary} output reminds us that posterior samples were not requested in the initial call to \code{spConjNNGP} or \code{summary}. Generating a reasonable number of posterior samples for parameters would take a few hours for a data set of this size.

\begin{knitrout}
\definecolor{shadecolor}{rgb}{0.969, 0.969, 0.969}\color{fgcolor}\begin{kframe}
\begin{alltt}
\hlstd{R> }\hlkwd{summary}\hlstd{(ser.c,} \hlkwc{digits} \hlstd{=} \hlnum{8}\hlstd{)}
\end{alltt}
\begin{verbatim}
## 
## Call:
## spConjNNGP(formula = ndvi ~ elev, data = ser.mod, coords = c("x", "y"), 
##     n.neighbors = 10, theta.alpha = theta.alpha, sigma.sq.IG = c(2, 
##         0.01), cov.model = "exponential", k.fold = 2, score.rule = "crps", 
##     X.0 = cbind(1, ser.ho$elev), coords.0 = as.matrix(ser.ho[, c("x", 
##         "y")]), n.omp.threads = 40)
## 
## Model class is NNGP, method conjugate, family gaussian.
## 
##             Estimate   Variance  
## (Intercept) 0.14544817 0.00002334
## elev        0.00155903 0.00000000
## sigma.sq    0.06520584 0.00000000
## phi         0.12000000 0.00000000
## alpha       0.01000000 0.00000000
\end{verbatim}

{\ttfamily\noindent\itshape\color{messagecolor}{\#\# If posterior summaries are desired, then either rerun spConjNNGP with\\fit.rep=TRUE, or specify the summary argument sub.sample to indicate the\\number of fitted and replicated samples to collect.}}\end{kframe}
\end{knitrout}

\begin{figure}[!htp]
  \centering
  \subfigure[]{\includegraphics[trim=0cm 2.5cm 0cm 2.5cm,clip,width=7cm]{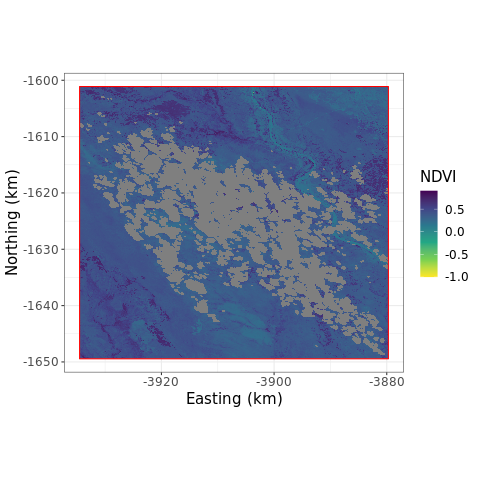}\label{kruger-ndvi-missing}}
  \subfigure[]{\includegraphics[trim=0cm 2.5cm 0cm 2.5cm,clip,width=7cm]{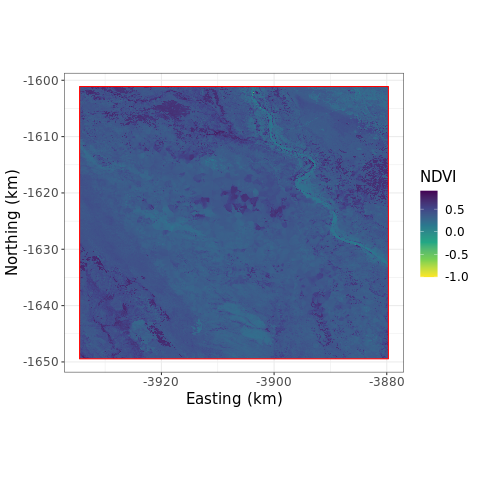}\label{kruger-yhat}}\\
  \subfigure[]{\includegraphics[trim=0cm 2.5cm 0cm 2.5cm,clip,width=7cm]{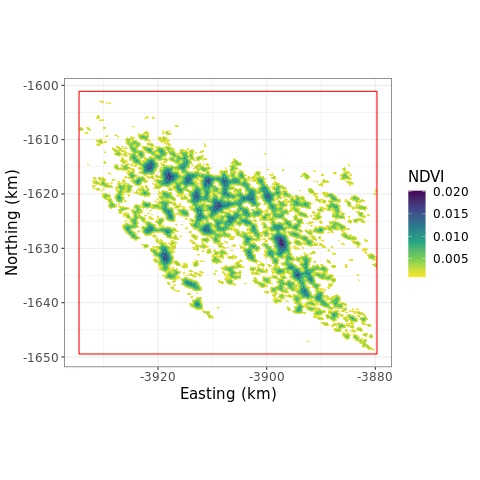}\label{kruger-yhatvar}}
  \caption{\subref{kruger-ndvi-missing} NDVI image subset identified by the red box in Figure~\ref{kruger-ndvi}, with gray pixels indicate missing NDVI data. \subref{kruger-yhat} missing pixel posterior predictive distribution mean. \subref{kruger-yhatvar} missing pixel posterior predictive distribution variance. }\label{krugerPred}
\end{figure}

For prediction, as illustrated in the previous analyses, we can pass the \code{spConjNNGP} object \code{ser.c} to \code{predict}. Alternatively, a slightly more efficient option (which avoids computing observed location covariance with their neighbor set twice, i.e., once for estimating parameters in \code{spCongNNGP} and again in subsequent call to \code{predict}) is to specify the prediction locations and associated design matrix via \code{coords.0} and \code{X.0} in the initial call to \code{spConjNNGP}. In this case, \code{ser.c} includes mean (\code{y.0.hat}) and variance (\code{y.0.hat.var}) estimates for the prediction locations. These predicted mean and variance of the mean estimates are shown for the missing pixels delineated by the red box in Figure~\ref{kruger-ndvi} (see Figure~\ref{kruger-ndvi-missing} for a zoomed in view) in Figures~\ref{kruger-yhat} and \subref{kruger-yhatvar}, respectively. As expected, the prediction variance surface shows that pixels close to observed pixels have higher precision due to borrowing of information through the spatial correlation structure.

\section{Summary} \label{sec:summary}

The \pkg{spNNGP} \proglang{R} package provides a suite of NNGP-based \citep{nngp} spatial regression models for both Gaussian and non-Gaussian point-referenced outcomes that are spatially indexed. The package implements the MCMC and MCMC-free algorithms detailed \cite{finley2019efficient} with the addition of the P{\'o}lya-Gamma latent variable model for binomial outcomes. Special care was taken to design algorithms that take advantage of multiprocessor computer via \proglang{OpenMP} and those with threaded BLAS and LAPACK libraries. Our future aim is to add  functionality to accommodate multivariate outcomes, where we envisage two settings, first, where a limited number of outcomes (e.g., fewer than 10) might be handled using a NNGP linear model of coregionalization \citep{Gelfand2004}, second, where the number of outcomes is larger and requires some dimension reduction, e.g., via a NNGP spatial factor model, akin to the model detailed in \cite{Taylor2019}. For such highly multivariate data, besides factor models, we will also explore new multivariate covariance functions using sparse inter-variable graphical models that parsimoniously capture the relationship between multiple variables. Future releases will also provide the flexibility to specify a general space-varying coefficient model like the \pkg{spBayes} \code{spSVC} function \citep{FinleyBanerjee2019}. Finally, our current functions for delivering exact Bayesian inference using conjugate models for the response will be expanded to accommodate latent spatial effects \citep{Banerjee2020} and multivariate spatial regression using the matrix-variate Normal and Inverse-Wishart families. 
\clearpage
\section*{Acknowledgments}
Finley was supported by National Science Foundation (NSF) EF-1253225 and DMS-1916395,and National Aeronautics and Space Administration's Carbon Monitoring System project. Banerjee was supported by NSF DMS-1513654, IIS-1562303, and DMS-1916349. Datta was supported by NSF DMS-1915803. The authors thank Michele Peruzzi for fruitful conversations about computing.

\bibliography{jss4016}

\end{document}